\documentclass[journal]{IEEEtran}
\usepackage{gensymb}
\usepackage{fancyhdr}
\usepackage{amsmath}
\usepackage{amssymb}
\usepackage{graphicx}
\usepackage{booktabs}
\usepackage{amsfonts}
\usepackage{multirow}
\usepackage{algorithm}
\usepackage{algorithmic}
\usepackage{mathrsfs}
\usepackage{bm}
\usepackage{exscale}
\usepackage{relsize}
\usepackage{setspace}
\usepackage{color}
\usepackage{comment}
\usepackage{ulem}
\usepackage{cite}
\usepackage{subfigure}

\newtheorem{lemma}{\textbf{Lemma}}
\newtheorem{theorem}{\textbf{Theorem}}
\newtheorem{remark}{Remark}

\DeclareMathOperator*{\argmax}{arg\,max}

\ifCLASSINFOpdf
\else
\fi
\begin{document}
%
\title{Enhanced Tracking and Beamforming Codebook Design for Wideband Terahertz Massive MIMO System}
%
%
%

	\author{
		Xu~Shi,~\IEEEmembership{Student~Member,~IEEE,}
		Jintao~Wang,~\IEEEmembership{Senior~Member,~IEEE,}
       Jian~Song,~\IEEEmembership{Fellow,~IEEE}
       
       	\thanks{
This work was supported in part by Tsinghua University-China Mobile Research Institute Joint Innovation Center. (Corresponding author: Jintao Wang.)

  	Xu Shi, Jintao Wang and Jian Song are with the Department of Electronic Engineering, Tsinghua University, Beijing 100084, China and Beijing National Research Center for Information Science and Technology (BNRist). (e-mail: shi-x19@mails.tsinghua.edu.cn; wangjintao@tsinghua.edu.cn; jsong@tsinghua.edu.cn).
}
	}

\maketitle

\begin{abstract}
True-time-delay (TTD) lines are recently applied inside Terahertz (THz) hybrid-precoding transceiver to acquire high beamforming gain against beam squint effect. However, beam tracking turns into a challenging puzzle where enormous potential beam directions bring about unacceptable overhead consumption. Frequency-scanning-based beam tracking is initially explored but still imperfect in previous studies. In this paper, based on TTD-aided hybrid precoding structure, we give an enhanced frequency-scanning-based tracking scheme. Multiple beams are generated and utilized simultaneously via several subcarriers for tracking at one timeslot. The squint beams' angular coverage at all subcarriers can be flexibly controlled by two different subcarrier-angular mapping policies, named forward-pairing and backward-pairing. Then multiple physical directions can be simultaneously searched in one timeslot for lower overhead consumption. Besides, closed-form searching radius bound, parameter configuration and interferences are theoretically analyzed. Furthermore, we provide the coupled codebook design for TTDs and phase shifters (PSs), with joint consideration of both beamforming and tracking. Analytical and numerical results demonstrate the superiority of the new frequency-scanning-based tracking scheme and beamforming codebook.

\end{abstract}


\IEEEpeerreviewmaketitle

\section{Introduction}

Exploitation of Terahertz (THz) spectrum is expected to accommodate enormous emerging wireless applications with high data throughput requirements \cite{THz1,THz2,THz3,THz4,THz5}. Benefited from the abundant band resources and high frequency,  the supported devices and transmission rate are prominently enlarged in THz wireless communication\cite{THz6}. To overcome the severe signal attenuation in THz band, massive multiple-input multiple-output (MIMO) is indispensable to generate narrow 'pencil' beam with enhanced array gain \cite{massiveMIMO1,massiveMIMO2}. Besides, given the limited hardware and power consumption in radio frequency (RF) chain, hybrid beamforming structure has been widely adopted in millimeter wave (mmWave) and THz systems to further upgrade energy efficiency \cite{hybridprecoding1,hybridprecoding2}.

Due to the massive number of antennas and huge spectrum bandwidth, beam squint effect is deadly harmful in wideband THz massive MIMO system \cite{squint1,squint2,squint5,squint_nonGao}. In beam squint effect, one propagation path points in distinct beam directions at different subcarriers since the shifted phase in realistic channel varies with frequency. Unfortunately, the conventional phase shifter (PS)-aided analog precoder  is frequency-independent and beam mismatch then appears in THz wideband beamforming, which may severely weaken communication quality. 

Up to now, the existing schemes against beam squint can be divided into two categories. One is based on the conventional PS-aided hybrid precoding structure, including window compensation \cite{beamforming_window} and beam broadening \cite{beamforming_broaden}. The former is to scale signals in spatial-frequency domain via energy-focusing window, and the latter is to establish several virtual sub-arrays for evenly distributed array gain across the whole frequency band. Other related hybrid precoding designs can also be found in \cite{beamforming_PS_fangjun,beamforming_PS_hanzo}. The other category is true time delay (TTD)-based \cite{beamforming_TTD_delay_phase,beamforming_TTD_2}. Inspired by large-bandwidth phased radar  system, we can perfectly mitigate beam squint by replacing PSs with TTDs, which can adaptively control the time delay of received signals from each antenna \cite{TTD1,TTD2}. Given that TTDs occupy  quite high hardware cost and power consumption, \cite{beamforming_TTD_delay_phase} established a hybrid network with a few TTDs and abundant PSs, which can approach near-optimal performance on sum-rate and array gain across the whole THz frequency band. All in all, the PS-aided schemes only partially relieve incompatibility inside beam squint, but fortunately the TTD-based method can be regarded as an exhaustively perfect enabler for beam squint compensation \cite{beamforming_broaden}.  

Besides, the channel state information (CSI) acquirement appears urgently challenging due to the lengthy pilot overhead as antennas grow in THz operating band \cite{THz4}. 
As antennas increase and beam width gets narrower, much more beams should be searched in a fixed angular interval for the optimal beamforming direction, which requires unaffordable pilot overhead due to the monogamy between timeslot and beam.
Therefore, traditional mmWave beam searching schemes are no longer appropriate  such as hierarchical sweeping \cite{tracking_hierarchical}, user mobility-assisted tracking \cite{tracking_mobility} like Kalman filtering \cite{spatial_rotation_UKF}, and so on \cite{tracking_Markov}. Fortunately, by adopting beam squint effect and TTD-aided beamforming structure, \cite{zooming_tracking} innovatively designed a specific pairing between frequency and beam angle, which is quite similar to the frequency-scanned radar.  Multiple beams are generated and utilized simultaneously via several subcarriers with only one RF chain for tracking, which also means that several beams can be searched at one timeslot and pilot overhead is then reduced by $90\%$.

However, there still exist some puzzles for TTD-aided THz wireless communications.  From point of view of tracking, limited by the beam pattern and pairing rule, the zooming pairing scheme \cite{zooming_tracking} only obtains good performance on semi-angular domain $[-1,0]$, but suffers from severe performance degradation when spatial direction is located inside $[0,1]$. And power leakage problem herein seems more challenging since the pairing for beam direction is non-uniform and previous fine-tune methods like spatial rotation \cite{spatial_rotation_UKF} cannot be employed directly. 
On the other hand, joint consideration of tracking and beamforming is necessary for TDD-aided system due to the same hardware support. Since TTDs and PSs are conjugate and both restricted to finite value sets in practice, quantized codebook design and resolution analysis are significant in TTD-PS-aided transceiver, especially to support both the beamforming and tracking modules simultaneously. Unfortunately, to our best knowledge, there exist few studies that thoroughly addressed the above puzzles. 

Correspondingly, the contributions of this article are summarized as follows: 

\begin{itemize}
\item 
Based on the TTD-aided frequency-dependent beams, we provide a new pairing policy between frequency and angular directions (called forward-backward-pairing), and further propose an enhanced wideband frequency-scanning-based tracking scheme. More efficient beam patterns can be generated to simultaneously search several beam angles via subcarrier indices at one timeslot, and the whole angular domain $[-1,1]$ can be robustly supported with lower pilot overhead.

\item We develop a novel compressive phase retrieval (CPR)-based method for power leakage compensation after wideband pairing-based tracking. In this gridless compensation scheme, we exploit the commonalities among multiple pilot timeslots and several subcarriers to further improve tracking accuracy. Simulation results confirm the superiority of the compensation scheme in wideband frequency-scanning-based tracking.

\item Besides, we further theoretically expand the tracking coverage under one timeslot for lower overhead. We derive closed-form upper bound of angular searching radius. In certain utmost directions, the frequency-dependent beam coverage, i.e., the angular tracking range, can be enlarged by $50\%$. 

\item Moreover, we design a quantized decoupled codebook for both TTDs and PSs in TDD-aided beamforming architecture, which is realistic and modified from the basic discrete Fourier transform (DFT) codebook. The new quantized codebook can simultaneously support TTD-aided beamforming and tracking with quite low degradation. The resolutions among codewords are then analyzed theoretically. 
\end{itemize}

	The rest of this paper is organized as follows. In Section \uppercase\expandafter{\romannumeral2}, we introduce the TTD-aided beamforming structure and analyze the propagation channel in time and frequency domain. In Section \uppercase\expandafter{\romannumeral3}, we propose the coupled codebook for TTDs and PSs in equivalent baseband model and further provide analysis for array gain and beam characteristics. The enhanced forward-backward wideband-mapping tracking algorithm is proposed in Section \uppercase\expandafter{\romannumeral4}, while we design a novel CPR-based energy-focusing scheme for power leakage problem. In Section \uppercase\expandafter{\romannumeral5} we give thorough analysis and evaluation for the proposed tracking scheme and obtain closed-form searching ranges. In Section \uppercase\expandafter{\romannumeral6} we provide the codebook resolution requirement with joint consideration of TTD-based beamforming and tracking. Finally, we show the simulation results in Section \uppercase\expandafter{\romannumeral7} and conclude this paper in Section \uppercase\expandafter{\romannumeral8}.

	\textit{Notations}: Lower-case and upper-case boldface letters $\mathbf{a}$ and $\mathbf{A}$ denote a vector and a matrix respectively. The operators $(\cdot)^*$, $(\cdot)^T$ and $(\cdot)^H$  denote conjugate, transposition and conjugate transposition of matrix. We use $(\cdot)^{-1}$ and $(\cdot)^{\dag}$ to denote inverse and pseudo-inverse. $\text{Tr}(\cdot)$ denotes matrix trace. $\otimes$ and $\circ$ denote the Kronecker product and Hadamard product, respectively. $\mathbf{I}_M$ is identity matrix with size $M\times M$. The operator $\text{diag}(\mathbf{a})$ is a square diagonal matrix with entries of $\mathbf{a}$ on its diagonal. $||\bm\Delta||_F$ and $||\bm\Delta||_2$ denote Frobenius norm and $l_2$ norm for matrix $\bm\Delta$. $\mathcal{CN}(\bm{\mu},\mathbf{\Sigma})$ stands for a circularly symmetric complex Gaussian distribution with mean $\bm{\mu}$ and variance $\mathbf{\Sigma}$.

\section{System Model}
Let us consider a wideband TTD-aided massive MIMO system as shown in Fig. 1(a), where the base station (BS) is configured with uniform linear array (ULA) of $N_\text{BS}$ antennas and $N_\text{RF}$ RF chains. The carrier frequency and system bandwidth are set as $f_c$ and $B$, respectively. Denote $d=\frac{c}{2f_c}$ as the fixed antenna spacing and $c$ is the propagation velocity of electromagnetic waves. For convenience, we assume the $2M+1$ baseband subcarriers are symmetric to zero with subcarrier spacing $f_d=\frac{B}{2M}$, which can be written as $f^\text{b}_{m}=mf_d,m=-M,\dots,M$. Furthermore, the up-conversion subcarrier frequencies can be written as
\begin{equation}
f_{m}=f_c+m f_d, \  m=-M,-M+1,\dots,M.
\end{equation}

In the TTD-aided hybrid beamforming transceiver as shown in Fig. 1(a),  $N_\text{TTD}$ TTD lines are equipped between each RF chain and corresponding $N_\text{BS}$ PSs. Specifically, overall $N_\text{BS}$ phase shifters assigned to each RF chain are divided into $N_\text{TTD}$ groups, i.e., the number of phase shifters assigned to each TTD is marked as $P$ and satisfies that $N_\text{TTD}\cdot P=N_\text{BS}$. According to \cite{beamforming_broaden}, the $k$-th TTD line's time-domain response can be written as $\delta(t-\hat{t}_k)$. Correspondingly, the frequency responses for PS and TTD  can be formulated as follows.
\begin{equation}
V_\text{PS}(\phi)=e^{-j\pi\phi}, \ \ V_\text{TTD}(\hat{t},f_m)=e^{-j2\pi f_m\hat{t}}=e^{-j\pi \frac{f_m}{f_c}t},
\end{equation}
where we define auxiliary time delay as
\begin{equation}
t\triangleq 2f_c \hat{t}
\end{equation}
and $\hat{t}$ is the real delay time caused by TTD.
By utilizing the TTD line network, different phase biases under several frequencies, caused by beam squint phenomenon, can be effectively compensated \footnote{The tradeoff for TTD number has been studied in \cite{beamforming_TTD_delay_phase} and near-optimal beamforming performance was theoretically confirmed with limited TTD lines.} .
And we can express the received signal at the subcarrier $f_m$ as:
\begin{equation}
\bm y_m=\mathbf{H}_m^H \mathbf{F}_m \mathbf{D}_m \bm s_m + \bm n_m,
\end{equation}
where $\mathbf{H}_m=[\mathbf{h}_{1,m},\dots,\mathbf{h}_{K,m}]\in \mathbb{C}^{N_\text{BS}\times K}$ represents THz channel matrix for $K$ users in the frequency $f_m$. $\mathbf{F}_m\in \mathbb{C}^{N_\text{BS}\times N_\text{RF}}$ and $\mathbf{D}_m\in \mathbb{C}^{N_\text{RF}\times N_\text{RF}}$ denote analog (PS+TTD) and baseband digital precoders, respectively. And $\mathbf{n} \sim \mathcal{CN}(\mathbf{0},\sigma_N^2\mathbf{I}_{K})$ represents additive white Gaussian noise (AWGN) at all $K$ users with noise power $\sigma_N^2$. The analog beamforming matrix $\mathbf{F}_m$ can be further decomposed as PS and TTD two parts as follows
\begin{equation}
	\arraycolsep=1.0pt\def\arraystretch{1.8}
	\begin{array}{lll}
\mathbf{F}_m&=&\mathbf{F}_\text{PS}(\bm\Phi) \circ \hat{\mathbf{F}}_{\text{TTD},m}(\bm T)\\
&=&\mathbf{F}_\text{PS}\circ (\mathbf{F}_{\text{TTD},m}\otimes \bm 1_{P\times 1}).
\end{array}
\end{equation}
where $\mathbf{F}_\text{PS}\in \mathbb{C}^{N_\text{BS}\times N_\text{RF}}$ and $\mathbf{F}_{\text{TTD},m}\in \mathbb{C}^{N_\text{TTD}\times N_\text{RF}}$ denote the frequency response matrix of phase shifters and TTD lines at $f_m$, respectively. $[\mathbf{F}_{\text{TTD},m}]_{n,k}=e^{-j\pi \frac{f_m}{f_c} t_{n,k}}$ denotes the response for the $n$-th TTD of $k$-th RF chain at subcarrier $f_m$, with auxiliary delay time set as $t_{n,k}$. Similarly, $[\mathbf{F}_\text{PS}]_{n,k}=e^{-j\pi\phi_{n,k}}$ is respective shifted phase through PS network.

\begin{figure}[!t]
			\includegraphics[width=0.9\linewidth]{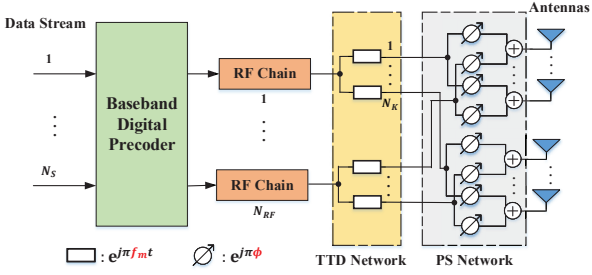}
			\label{system_model_a}

		\caption{Block diagram of TTD-aided beamforming model.}
\end{figure}

The widely utilized ray-based time-domain channel  impulse response \cite{THzChannel} between the $k$-th UE and the $n$-th wideband THz BS antenna $h_{k,n}(t)$ is expressed as
\begin{equation}
h_{k,n}(t)=\sum_{l=1}^L g_{k,l} e^{-j2\pi(n-1)\psi_{k,l}}\delta\left(t-\tau_{k,l}^0-(n-1)\cdot\frac{\psi_{k,l}}{f_c}\right),
\end{equation}
where $\delta(\cdot)$ denotes the Dirac delta function, $\psi_{k,l}$ is the $l$-th beam angle of arrival (AoA) and $(n-1)\frac{\psi}{f_c}$ is the delay difference among BS antennas which causes the spatial-wideband effect \cite{squint5}.  By transforming the impulse response into frequency domain and reshaping it into a vector along all antennas,  we can obtain the corresponding frequency response:
\begin{equation}
\mathbf{h}_{k,m}=\sum_{l=1}^L g_{k,l}e^{-j2\pi f_m^b \tau_{k,l}^0}\mathbf{a}_{N_\text{BS}}(f_m,\psi_{k,l}).
\end{equation}
and the frequency-dependent steering vector
\begin{equation}
\mathbf{a}_N(f_m,\psi)\triangleq  [1,e^{-j\pi \frac{f_m}{f_c}\psi},\dots,e^{-j\pi \frac{f_m}{f_c}(N-1)\psi}]^T.
\label{channel_frequency}
\end{equation}

Due to the severe path loss under extreme high frequency and large antenna number, line-of-sight (LoS) beam is dominant among all channel paths in THz communication. Besides, it is worth noting that we mainly focus on beam tracking and beamforming codebook design of wideband THz system. Without loss of generality, we only consider one UE (one RF chain) scenario in this paper \footnote{Multi-user scenario can be similarly extended with additional consideration of  Inter-User Interference (IUI) in beam selection module, which has been deeply studied in conventional mmWave hybrid beamforming.}.  Then the corresponding simplified signal in $l$-th timeslot can be rewritten as:
\begin{equation}
y_{m,l}=\mathbf{h}_m^H\left[\mathbf{f}_\text{PS}(\bm \phi_l)\circ (\mathbf{f}_{\text{TTD},m}(\bm t_l)\otimes \bm 1_{P\times 1})\right]s_l+n_m.
\end{equation}
where $\bm \phi\in \mathbb{C}^{N_\text{BS}\times 1}$ is phase shift vector and $\bm t\in \mathbb{C}^{N_\text{TTD}\times 1}$ is the time delay vector.

\section{Codebook Design for TTD-aided Model}
In this section we give the codebook design for TTD-aided system, which is the basis for all practical beamforming and tracking modules next.  Moreover, it is necessary to analyze the codebook-based array gain and corresponding beam characteristics in this section, since the analysis and inference are extremely instructive for the following beam tracking, codebook design and performance analysis. 

Inspired by the derivation from \cite{beamforming_TTD_2,zooming_tracking}, we establish an equivalent beamforming model as shown in Fig. 1(b). We divide the TTD phase $e^{-j\pi \frac{f_m}{f_c} t_n}$ into frequency-independent part $e^{-j\pi t_n}$ and baseband part $e^{-j\pi \frac{f_m^b}{f_c} t_n}$, and merge the former part into PS network. Define the equivalent PS values as $\bm\psi=\bm\phi+(\bm t\otimes 1_{P\times 1})$,  and the equivalent signal model \footnote{Notice that the tuple $\{\bm f_\text{PS}(\bm\psi),f_m^b(\bm t)\} $ is a bijection to $\{ \bm f_\text{PS}(\bm \phi), \bm f_m(\bm t) \}$ and is reversible from the latter, which also means the two models are equivalent.} can be further rewritten as 
\begin{equation}
y_m=\mathbf{h}_m^H\left[\mathbf{f}_\text{PS}(\bm \psi)\circ (\mathbf{f}_{\text{TTD},m}^b(\bm t)\otimes \bm 1_{P\times 1})\right]s+n_m,
\end{equation}
where baseband time delay $\bm f_m^b(\bm t)$ is defined as $[\bm f_m^b(\bm t)]_{n}=e^{-j\pi \frac{f_m^b}{f_c} t_n}$. Intuitively speaking, the equivalent PS array $\mathbf{f}_\text{PS}$ provides the rough beam direction while the baseband TTD array $\mathbf{f}^b_{\text{TTD},m}$  undertakes a fine-tune for compensation of beam squint. Therefore, the codebook for $\bm\psi$ and $\bm t$ can be designed as discrete Fourier transform (DFT)-style:
\begin{equation}
	\arraycolsep=1.0pt\def\arraystretch{1.8}
	\begin{array}{lll}
	\bm t=P[0,t,\dots,(N_\text{TTD}-1)t]^T&,& 
	\\
\bm \psi= [0,\psi,\dots,(N_\text{BS}-1)\psi]^T&, & \psi\in \mathbb{U}^Q_{R_{\psi}}([-1,1])\\
\end{array}
\label{codeword}
\end{equation}
where $\mathbb{U}^Q_R([a,b])$ denotes uniform quantization for interval $[a,b]$ with sampling spacing (resolution) $R$. After the codeword selection of $\bm\psi$ and $\bm t$, we can directly calculate realistic PS value $\bm\phi$ and delay time $ \hat{\bm t}$ to further control the initial TTD-aided transceiver. The quantization resolution design of $t$ and $\psi$ in (\ref{codeword}) will be shown in Section VI, with joint consideration of beamforming and tracking. So for brevity, we don't illustrate too much about codebook performance in this section.

\begin{remark}
The equivalent signal model and corresponding codebook generation for TTD-aided beamforming simultaneously support wideband beamforming and tracking. To the best of our knowledge, it is first clearly proposed and jointly considered from the two aspects in this paper, which is necessary and useful, although some brilliant studies have been done from partial aspects such as \cite{beamforming_TTD_2,zooming_tracking}. 
\end{remark}

Next, we analyze the array gain  based on the codebook above, which can be calculated as follows:
\begin{equation}
\arraycolsep=0.5pt\def\arraystretch{2.5}
	\begin{array}{lll}
     &\displaystyle \mathcal{G}(f_m,\theta) = \left|\frac{1}{N_\text{BS}}\mathbf{a}_N(f_m,\theta)^H\mathbf{f}_m\right|&\\
     =&\displaystyle \frac{1}{N_\text{BS}}\left|\sum_{p=1}^{P}\sum_{q=1}^{N_\text{TTD}}e^{j\pi \left[(q-1)P+(p-1)\right]\frac{f_m}{f_c}\theta}\times\right.&\\
    &\displaystyle\ \ \ \ \ \ \ \ \ \ \ \ \ \ \ \ \left. e^{-j\pi[(q-1)P+(p-1)]\psi}\times e^{-j\pi (q-1)\frac{f_m^b}{f_c}Pt}\right|&\\
    =&\displaystyle \frac{1}{N_\text{BS}}\left|\sum_{p=1}^{P}e^{j\pi (p-1)(\frac{f_m}{f_c}\theta-\psi)}\sum_{q=1}^{N_\text{TTD}}e^{j\pi P(q-1)(\frac{f_m}{f_c}\theta-\psi-\frac{f_m^b}{f_c}t)}\right|&\\
    =&\displaystyle \frac{1}{P N_\text{TTD}}\left|\frac{\sin\frac{P\pi}{2}(\frac{f_m}{f_c}\theta-\psi) }{\sin\frac{\pi}{2}(\frac{f_m}{f_c}\theta-\psi)}\right|\cdot \left| \frac{\sin \frac{N_\text{BS}\pi}{2}(\frac{f_m}{f_c}\theta-\psi-\frac{f_m^b}{f_c}t)}{\sin \frac{P\pi}{2}(\frac{f_m}{f_c}\theta-\psi-\frac{f_m^b}{f_c}t)} \right|&\\
    =& \displaystyle \Xi_P\left(\frac{f_m}{f_c}\theta-\psi\right) \cdot \Xi_{N_\text{TTD}}\left(P\left(\frac{f_m}{f_c}\theta-\psi-\frac{f_m^b}{f_c}t\right)\right)&
    \end{array}
    \label{gain_formulation}
\end{equation}
where 
\begin{equation}
\Xi_N(a)\triangleq \frac{1}{N}\left|\frac{\sin \frac{N\pi}{2}a}{\sin \frac{\pi}{2}a}\right|
\end{equation}
is the normalized Dirichlet function, with period $T_\text{Dir}=2$ and main lobe's semi-width $W_\text{Dir}^\text{semi}=\frac{2}{N}$ \cite{Diriclet}.

\begin{figure}[!t]
\centering
\includegraphics[width=0.8\linewidth]{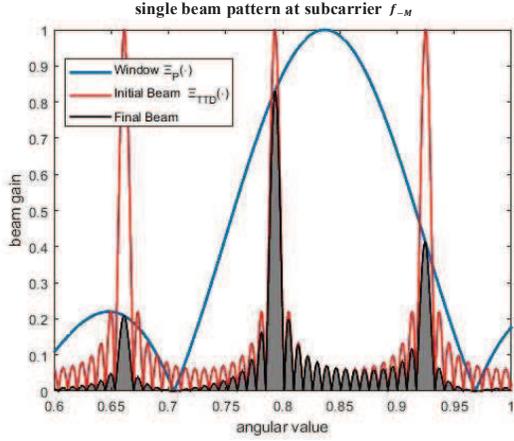}
\caption{The beam patterns of $\Xi_P(\cdot)$, $\Xi_{N_\text{TTD}}(\cdot)$ and final beam $\mathcal{G}$ under single subcarrier $f_{-M}$.}
\label{diric_function}
\end{figure}

Based on the derivation in (\ref{gain_formulation}), we exploit some interesting properties of the two Dirichlet functions $\Xi_P(\cdot)$ and $\Xi_{N_\text{TTD}}(\cdot)$, which may be helpful for the following beam tracking and performance analysis. We herein regard real channel angle $\theta$ as the function argument. As for the former $\Xi_P(\frac{f_m}{f_c}\theta-\psi)$, the period, semi-width and the global-maximum point location are calculated as:
\begin{equation}
\arraycolsep=0.5pt\def\arraystretch{1.8}
	\begin{array}{ccc}
&&\displaystyle T^{(1)}_m=\frac{2f_c}{f_m}, \ \ W^{(1)}_m=\frac{2f_c}{Pf_m},\ \ \displaystyle\theta^{(1)}_{c,m}=\frac{f_c}{f_m}\psi
\end{array}
\label{outer_window_property}
\end{equation}
Similarly, as for the latter $\Xi_{N_\text{TTD}}(P(\frac{f_m}{f_c}\theta-\psi-\frac{f_m^b}{f_c}t))$, the period, semi-width and the global-maximum point location are
\begin{equation}
\arraycolsep=0.5pt\def\arraystretch{1.8}
	\begin{array}{lll}
&&\displaystyle T^{(2)}_m=\frac{2f_c}{Pf_m}, \ \ W^{(2)}_m=\frac{2f_c}{N_\text{BS}f_m},\ \ \displaystyle\theta^{(2)}_{c,m}=\frac{f_c}{f_m}\psi+\frac{f_m^b}{f_m}t
\end{array}
\label{inner_beam_property}
\end{equation}

It is worth noting that the former period $T^{(1)}_m$ is much larger than $T^{(2)}_m$ and near to the width of argument interval $[-1,1]$. Therefore, we can regard the former Dirichlet function $\Xi_P(\frac{f_m}{f_c}\theta-\psi)$ as a window for the second term $\Xi_{N_\text{TTD}}(\cdot)$ and almost all components outside the main lobe of $\Xi_P$ are filtered to zero, as depicted in Fig. 2. Besides, notice that the former period $T^{(1)}_m$ is equal to the latter semi-width $W^{(2)}_m$, which means that there must exist only two peaks inside main lobe of $\Xi_P$ (when $\theta^{(1)}_{c,m}\neq\theta^{(2)}_{c,m}$). The maximum array gain at subcarrier $f_m$ almost always appears at $\theta=\theta_{m,c}^{(2)}+kT_m^{(2)}$ and must have $|\theta-\theta_{c,m}^{(1)}|<W_m^{(1)}$. 

For \emph{wideband beamforming}, we set PS and TTD parameters equal to the physical user direction $\theta_0$ ideally, i.e., $\psi=t=\theta_0$. In this way, the global-maximum point for array gain $\mathcal{G}$ locates at $\theta=\theta_0$ for all subcarriers $f_m$, which means beam squint is perfectly eliminated via TTD line. 

As for TTD-aided \emph{wideband tracking}, we configure the equivalent PS/TTD values $\psi, t$  bias to real beam angle. Notice the maximum array gain at subcarrier $f_m$ must lie in $\theta^{(2)}_{m,c}\ (+kT_m^{(2)})$ and thus we can strictly design the values of $\psi$ and $t$ to form particular beam patterns for tracking, i.e., establish a monotonical bijection between subcarriers $\{f_m\}$ and quantized angle set $\{\theta_m\}$ via maximum array gain location (\ref{inner_beam_property}), i.e.,
\begin{equation}
\theta_m=\frac{f_c}{f_m}\psi+\frac{f_m^b}{f_m}t,\  m=-M,\dots,M.
\label{16}
\end{equation}
where $\theta_m$ denotes the angular location with maximum beamforming gain at frequency $f_m$. 

For clearer illustration of the frequency-scanning tracking, we provide the holistic gain distribution of overall subcarriers in Fig. 4(b), where $2M+1$ beamforming curves are plotted, corresponding to the whole $2M+1$ subcarriers. Each curve here displays a similar beam pattern to the gray region of Fig. 2. We can obviously observe the monotonical bijective relationship between subcarriers $\{f_m\}$ and quantized angular values $\{\theta_m\}$ in Fig. 4(b). Following this mechanism, we can easily screen the maximum beamforming gain, corresponding subcarrier and angular direction to implement tracking procedure in one timeslot.

\section{Enhanced Tracking}

\subsection{Drawback of previous frequency-scanned tracking}
Before we propose the enhanced tracking scheme, it is necessary to illustrate the conventional zooming-based method \cite{zooming_tracking} and its drawback in some specific scenarios. 
Here we only consider one timeslot for tracking, i.e., we only send pilot $s_l$ once, with $L=1$.  
Assume the initial beam searching range is set as interval $[\theta_0-\alpha,\theta_0+\alpha]$, where $\theta_0$ denotes the central angle and $2\alpha$ is the whole searching length in angular domain.  Based on the TTD array gain derived in (\ref{gain_formulation}), \cite{zooming_tracking} projects all subcarriers to beam angles forwardly as shown in Fig. \ref{forward_backward}(a). Fix the peripheral mapping $f_{-M} \rightarrow \theta_0-\alpha$ and $f_{M}\rightarrow \theta_0+\alpha$ and we have
\begin{equation}
\left\{
    \arraycolsep=0.5pt\def\arraystretch{1.8}
        \begin{array}{ccc}
        \theta_0-\alpha&=&\displaystyle\frac{f_c}{f_c-Mf_{d}}\psi^\text{fw}-\frac{Mf_{d}}{f_{c}-Mf_d}t^\text{fw}\\
            \theta_0+\alpha&=&\displaystyle\frac{f_c}{f_c+Mf_{d}}\psi^\text{fw}+\frac{Mf_{d}}{f_c+Mf_{d}}t^\text{fw}\\
        \end{array}
\right. ,
\label{forward_relation}
\end{equation}
and the final configuration of equivalent PS/TTD can be directly calculated via (\ref{forward_relation}):
\begin{equation}
\left\{
    \arraycolsep=0.5pt\def\arraystretch{1.8}
        \begin{array}{ccc}
        \psi^\text{fw} &=&\displaystyle\theta_0+\frac{Mf_d}{f_c}\alpha\\
            t^\text{fw}&=&\displaystyle \theta_0+\frac{f_c}{Mf_{d}}\alpha
        \end{array}
\right. .
\label{forward_codebook_setting}
\end{equation}
Then frequency-dependent beams can be further generated to simultaneously track several physical directions, i.e., the angle-domain coverage $[\theta_0-\alpha,\theta_0+\alpha]$, as depicted in Fig. \ref{beam_diffusion}. 


However, the coverage radius $\alpha$ here cannot be arbitrarily configured but should satisfy certain strict constraints. 
To avoid passive influence of sidelobes inside $\Xi_{N_\text{TTD}}(\cdot)$ and periodic beams inside $\Xi_P(\cdot)$, we should ensure that $\theta_{c,m}^{(2)}$ is located around the strong mainlobe of windowing function $\Xi_P(\cdot)$ with maximum angular radius $W_m^{(1)}/2$, i.e., 
\begin{equation}
\theta_{c,m}^{(2)}\in \left[\theta_{c,m}^{(1)}-\frac{W_m^{(1)}}{2}, \theta_{c,m}^{(1)}+\frac{W_m^{(1)}}{2} \right], \forall m.
\label{forward_constraint}
\end{equation}
Substituting (\ref{outer_window_property}), (\ref{inner_beam_property}) and (\ref{forward_codebook_setting}) into (\ref{forward_constraint}), then the constraint for $\alpha$ is simplified as follows
\begin{equation}
\alpha^{\text{fw}}<\frac{1}{P}-\frac{Mf_d}{f_c}\theta_0.
\label{forward_range}
\end{equation}
When we further enlarge the range $\alpha>\frac{1}{P}-\frac{M f_d}{f_c}\theta_0$, there appears a diffusion in the multi-frequency beam pattern as shown in Fig. \ref{beam_diffusion}(a), which causes a mismatch between subcarriers and quantized angles and thoroughly wrecks the zooming-based tracking. Furthermore, from (\ref{forward_range}) we can easily observe the drawbacks of conventional zooming-based method (we only consider interval $\theta_0\in [0,1]$ here):
\begin{itemize}
\item Tracking fails when $\theta_0\rightarrow 1$: 
Actually, the upper bound $\frac{1}{P}$ can only be approached when central angle $\theta_0=0$. When UE moves to certain specific directions that $\theta_0\approx  1$, the maximum searching range $\alpha$ declines to $\frac{1}{P}-\frac{M f_d}{f_c}$, which may be quite small and consume unaffordable time resource for tracking. For example, when we set $P=16$, $f_c=100\ \text{GHz}$, $B=12.5\ \text{GHz}$ and AoD $72\degree$ ($\theta_0=0.95$), the corresponding searching radius $\alpha|_{\theta_0=0.95}\approx 3\times 10^{-3}$. Almost $34$ pilot timeslots are needed to thoroughly scan the angular interval $[0.9,1]$. 

\item Angle-dependent range $\alpha$: On the other hand, we cannot find a constant scanning range for all directions, because $\alpha$ is angle-dependent here and the minimum searching radius $\frac{1}{P}-\frac{M f_d}{f_c}$ is too small for pilot consumption. The adaptive adjustment for $\alpha$ is redundant and hardware-intricate during tracking process. 

\item Power leakage problem: The zooming-based tracking \cite{zooming_tracking} still only searched part quantized angular values, which causes beam mismatch since there may exist biases between realistic physical direction and tracked angle. Therefore, it is necessary to design enhanced frequency-scanning-based tracking scheme with enlarged and fixed searching range $\alpha$, also with consideration of power leakage compensation. 
\end{itemize}

\subsection{Enhanced Tracking Design}

\begin{figure}[!t]
\centering
\includegraphics[width=1\linewidth]{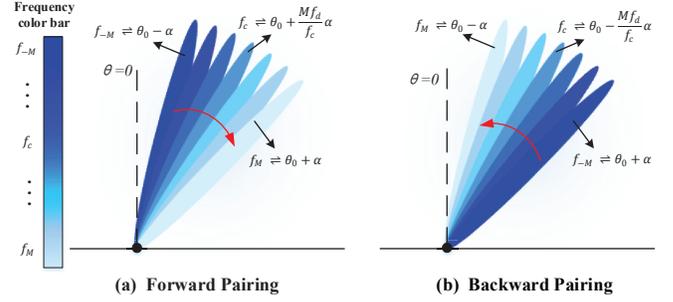}
\caption{The tracking procedures of forward and backward pairings between subcarriers and angular values.}
\label{forward_backward}
\end{figure}

\begin{figure*}[!t]
\centering
\includegraphics[width=0.8\linewidth]{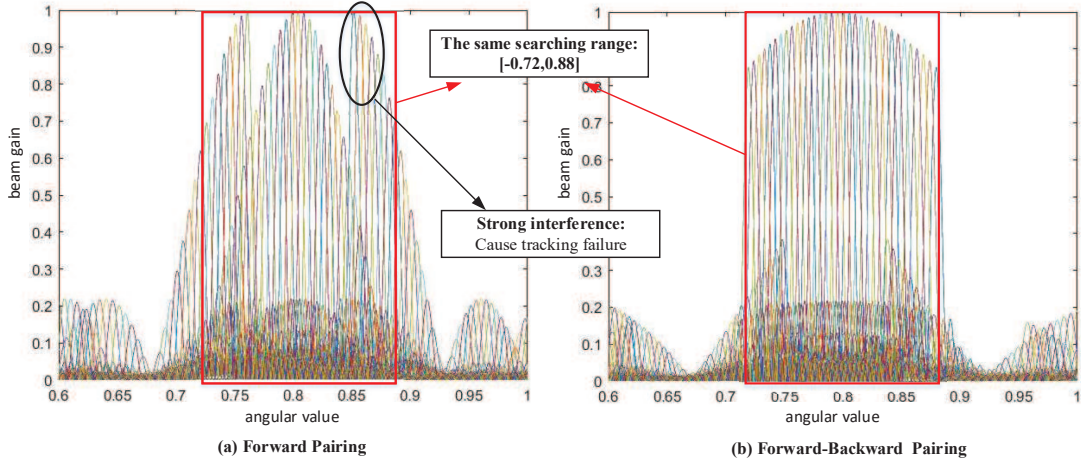}
\caption{The beam pattern comparison of conventional forward mapping and forward-backward pairing schemes with overall $128$ subcarriers, where the central angle $\theta_0=0.8$ and searching radius $\alpha=0.08$.}
\label{beam_diffusion}
\end{figure*}

To overcome the disadvantages analyzed above, we propose a novel wideband THz tracking scheme named \emph{Forward-Backward-pairing} based algorithm.  

As illustrated in Section III, we should establish a bijection between subcarriers $\{f_m\}$ and angular searching set $\{\theta_m\}\subset [\theta_0-\alpha,\theta_0+\alpha]$, where the sampled angular points satisfy
\begin{equation}
\theta_0-\alpha=\theta_{-M}<\dots<\theta_m<\dots<\theta_{M}=\theta_0+\alpha.
\end{equation}
The conventional zooming tracking only considered forward projection $f_m \rightarrow \theta_m$. However, there exists another projection approach named as backward mapping, i.e., $f_m \rightarrow \theta_{-m}$ as shown in Fig. \ref{forward_backward}(b). By jointly adopting the forward and backward mappings in different scenarios, we can give a better frequency-scanning-based tracking scheme. 

First, let's provide some analysis about the backward mapping similarly to (\ref{forward_relation}). The peripheral backward mappings $f_{-M}\rightarrow \theta_M$ and $f_M\rightarrow \theta_{-M}$ can be established as 
\begin{equation}
\left\{
    \arraycolsep=0.5pt\def\arraystretch{1.8}
        \begin{array}{ccc}
        \theta_0-\alpha&=&\displaystyle\frac{f_c}{f_c+Mf_{d}}\psi^\text{bw}+\frac{Mf_{d}}{f_{c}+Mf_d}t^\text{bw}\\
            \theta_0+\alpha&=&\displaystyle\frac{f_c}{f_c-Mf_{d}}\psi^\text{bw}-\frac{Mf_{d}}{f_c-Mf_{d}}t^\text{bw}\\
        \end{array}
\right. ,
\label{backward_relation}
\end{equation}
and based on (\ref{backward_relation}), the configuration $\{\psi,t\}$ is solved as 
\begin{equation}
\left\{
    \arraycolsep=0.5pt\def\arraystretch{1.8}
        \begin{array}{ccc}
        \psi^\text{bw} &=&\displaystyle\theta_0-\frac{Mf_d}{f_c}\alpha\\
            t^\text{bw}&=&\displaystyle \theta_0-\frac{f_c}{Mf_{d}}\alpha
        \end{array}
\right. .
\label{backward_codebook_setting}
\end{equation}
Substituting (\ref{backward_codebook_setting}) into (\ref{forward_constraint}), the maximum angular searching range is enlarged to
\begin{equation}
\alpha^\text{bw}<\frac{1}{P}+\frac{Mf_d}{f_c}\theta_0.
\end{equation}

Compared with the forward-pairing range $\alpha^\text{fw}$, we can easily observe that  as central angle $\theta_0$ increase, forward range $\alpha^\text{fw}$ drop off while backward range $\alpha^\text{bw}$ rise up. And they  intersect at $\theta_0=0$, which inspires us to retain the maximum value inside $\alpha^\text{bw}$ and $\alpha^\text{fw}$ as
\begin{equation}
    \arraycolsep=0.5pt\def\arraystretch{1.8}
        \begin{array}{lll}
\alpha^\text{b-f}&=&\max \{\alpha^\text{bw},\alpha^\text{fw} \}=\left\{
        \begin{array}{cc}
\alpha^\text{fw},&\ \theta_0<0 \\
\alpha^\text{bw},& \ \theta_0\geq 0
        \end{array}
        \right. \\
        &<&\displaystyle \frac{1}{P}+\frac{Mf_d}{f_c}|\theta_0|.
        \end{array}
        \label{F_B_pairing_range}
\end{equation}

When tracking UEs with angle range $[\theta_0-\alpha,\theta_0+\alpha]$, to pursue lower pilot overhead for UE tracking and approach the bound $\bar{\alpha}^\text{b,f}= \frac{1}{P}+\frac{Mf_d}{f_c}|\theta_0|$, we divide it into two categories $\theta_0<0$ and $\theta_0\geq 0$. If the central angle $\theta_0$ is a negative number, we adopt the forward mapping $f_m\rightarrow \theta_m$ and then the corresponding tracking process is the same with \cite{zooming_tracking}, as formulated in (\ref{forward_range}).  On the contrary, if $\theta_0\geq 0$, we utilize the backward mapping $f_m\rightarrow \theta_{-m}$ instead of forward one.  

Generally speaking,  we now assume there exist $L$ timeslots for tracking, and only uniformly-quantized angular fraction $[\theta_0-\alpha+\frac{2\alpha}{L}(l-1),\ \theta_0-\alpha+\frac{2\alpha}{L}l]$ is tracked  in the $l$-th timeslot. After we set up the subcarrier-angle mapping configuration (\ref{F_B_pairing_range}), i.e., the parameters $\psi_l$ and $t_l$, the signal processing for enhanced TTD-aided tracking is then shown as follows. The received signal at $l$-th frame is formulated as 
\begin{equation}
y_{l,m} = \mathbf{h}_{m}^H\mathbf{f}_{l,m}(\psi^{\text{b-f}}_l,t^{\text{b-f}}_l)s_{l,m}+n_{l,m}.
\label{signal_model_tracking}
\end{equation}
Without loss of generality, the pilot sequences $s_{l,m}$ of all $L$ timeslots are fixed as $1$. Then the strongest beam and corresponding timeslot index triple $\{m,l\}$ can be selected from overall received signals $\mathbf{Y}\in \mathbb{C}^{L\times (2M+1)}$:
\begin{equation}
\{\hat{m},\hat{l}\}=\argmax_{l=1,\dots,L, \ m=-M,\dots,M} \|[\mathbf{Y}]_{l,m}\|^2.
\label{selection_strongest}
\end{equation}
The final beam angle can be calculated as
\begin{equation}
\hat{\theta} =\displaystyle \frac{f_c}{f_{\hat{m}}}\psi_{\hat{l}}+\frac{f^b_{\hat{m}}}{f_{\hat{m}}}t_{\hat{l}},
\label{tracked_angle}
\end{equation}
where $\psi_{\hat{l}}, t_{\hat{l}}$ are controlled by $\hat{l}$-th time fraction's central angle $\theta_{0}^{\hat{l}}\triangleq \theta_0-\alpha+\frac{(2\hat{l}-1)\alpha}{L}$ and tracking radius $\frac{\alpha}{L}$. The detailed process is provided in Algorithm 1.

	\begin{algorithm}[htb] 
		\normalem
		\caption{Proposed Wideband THz Tracking:  Forward-Backward-Pairing Tracking  } 
		\label{alg1} 
		\begin{algorithmic}[1] 
			\REQUIRE Central direction for searching $\theta_0$; Searching radius $\alpha$; Pilot overhead length $L$; Transceiver configuration $N_\text{BS}$, $N_\text{TTD}$, $P$; Frequency info $f_c$, $f_d$ and $M$. 

			\ENSURE Physical beam angle $\hat{\theta}$
			
			
			\FOR{ time fraction index $l=1$ to $L$}
            
			\STATE fraction info $\theta_{0,l}\leftarrow \theta_0-\alpha+\frac{(2l-1)\alpha}{L}$; $\alpha_l\leftarrow \alpha/L$
            
            \IF{$\theta_{0,l}\geq 0$}
            \STATE Backward pairing: calculate $\psi_l,\  t_l$ via (\ref{backward_codebook_setting})
            \ELSE
            \STATE Forward pairing: calculate $\psi_l,\ t_l$ via (\ref{forward_codebook_setting})
            \ENDIF
            \STATE calculate precoding matrix $\mathbf{f}_{l,m}(\psi_l,t_l)$ via (\ref{codeword})
            \STATE obtain received signals $\mathbf{y}_l=[y_{l,-M},\dots,y_{l,M}]^T$ via (\ref{signal_model_tracking})
            \ENDFOR
            
            \STATE pick the strongest beam and corresponding subcarrier/time fraction index $\{\hat{l},\hat{m}\}$ from all received signals $\mathbf{Y}$ via (\ref{selection_strongest})
            \STATE calculate the final physical beam angle $\hat{\theta}$ via (\ref{tracked_angle})
		\end{algorithmic}
	\end{algorithm}

\subsection{Power Leakage Elimination}

Although the forward-backward-pairing tracking can significantly reduce pilot overhead consumption, there still exists power leakage problem since subcarriers for direction mapping are finite. When the real physical direction $\theta^\text{real}$ don't match the quantized searching grid $\hat{\theta}\in\{\theta_{l,m}\}$ exactly,  power leakage effect then appears. Under certain extreme situations such as when massive antennas are configured, the beamforming mainlobe amplitude is sharply reduced and side-lobe interference increases, which is quite harmful to communication capacity and beamforming gain. To tackle this problem, based on our proposed TTD-aided tracking above, we further provide a CPR-based scheme for power leakage compensation.

Based on the channel model (\ref{channel_frequency}) and only considering the dominant LoS path, the initial objective can be established as follows,
\begin{equation}
    \arraycolsep=0.5pt\def\arraystretch{1.8}
        \begin{array}{lll}
\hat{\theta}^\text{P}&=&\displaystyle\mathop{\arg\min}\limits_{\theta} \ \sum_{l=1}^L \sum_{m=-M}^{M}\left|y_{l,m}-\mathbf{f}_{l,m}^H\mathbf{h}_m(\theta)\right|^2\\
&=&\displaystyle \mathop{\arg\min}\limits_{\theta} \ \sum_{l=1}^L \sum_{m=-M}^{M}\left|y_{l,m}-g_m\mathbf{f}_{l,m}^H\mathbf{a}_m(\theta)\right|^2\\
\end{array}
\label{initial_objective}
\end{equation}
where $g_m$ denotes the complex gains for LoS path at frequency $f_m$. Notice that the path gain's amplitude $|g_m|$ is invariable  under all subcarriers due to the only LoS path, which means we can divide it into two separate parts $g_m=g e^{j\tau_m}$. Besides, the path gain's phase $\tau_m$ is irrelevant to timeslot fraction index $l$, and thus we can merge all timeslots and further reformulate the objective (\ref{initial_objective}) into
\begin{equation}
\min_{\theta,g,\tau_m} \ \ \epsilon = \sum_{m=-M}^{M} \left|\hat{\mathbf{y}}_m- g e^{j\tau_m} \mathbf{B}_m^H \mathbf{a}_m(\theta)   \right|^2,
\label{further_objective}
\end{equation}
where $\hat{\mathbf{y}}_m=[y_{1,m},y_{2,m},\dots,y_{L,m}]^T$ and $\mathbf{B}_m=[\mathbf{f}_{1,m},\mathbf{f}_{2,m},\dots,\mathbf{f}_{L,m}]$  denote received signals and joint beamforming matrix along all timeslots $L$ for the $m$-th subcarrier, respectively.

It is difficult to directly calculate the global optimum for problem (\ref{further_objective}) due to its non-convexity. Therefore, we adopt alternating minimization method searching for the near-optimal solution. Take the result from Algorithm 1 as an initial value, marked as $0$-th iteration's result $\hat{\theta}^{(0)}$. During each iteration, we optimize beam gain amplitude $g$, phase $e^{j\tau_m}$ and beam direction $\theta$ successively.

First, we neglect the phase information and only consider the modulus to optimize the common gain amplitude $g$. The problem is derived as follows:
\begin{equation}
\min_{g^{(t+1)}} \ \sum_{m=-M}^{M} \left| |\hat{\mathbf{y}}_m|-g|\mathbf{B}_m^H\mathbf{a}_m(\hat{\theta}^{(t)})|  \right|^2,
\end{equation}
which can be easily calculated via least square (LS) as
\begin{equation}
g^{(t+1)}_\text{opt}=\frac{\displaystyle\sum_{m=-M}^{M} \left[ 2|\hat{\mathbf{y}}_m^H|\cdot|\mathbf{B}_m^H\mathbf{a}_m(\hat{\theta}^{(t)})| \right]}{\displaystyle\sum_{m=-M}^{M} \left[  2\mathbf{B}_m^H\mathbf{a}_m(\hat{\theta}^{(t)})\mathbf{a}_m(\hat{\theta}^{(t)})\mathbf{B}_m \right]}
\label{gain_opt}
\end{equation}

Next, we optimize the phase parameters separately for all subcarriers $m=-M,\dots,M$ from (\ref{further_objective}): 
\begin{equation}
\tau_{m,\text{opt}}^{(t+1)}=\text{angle}\left(\frac{\displaystyle \mathbf{a}_m(\hat{\theta}^{(t)})^H\mathbf{B}_m\hat{\mathbf{y}}_m}{\displaystyle |\mathbf{a}_m(\hat{\theta}^{(t)})^H\mathbf{B}_m\hat{\mathbf{y}}_m|}\right)
\label{phase_opt}
\end{equation}

\newcounter{mytempeqncnt}
\begin{figure*}[!t]
	\normalsize
	\setcounter{mytempeqncnt}{\value{equation}}
	\setcounter{equation}{33}
	\begin{equation}
    \arraycolsep=0.5pt\def\arraystretch{2.5}
        \begin{array}{lll}
 & \nabla \epsilon \left.\right|_\theta =\displaystyle \sum_{m=-M}^M  \left[ \left(\hat{\mathbf{y}}_m- g^{(t+1)}_\text{opt}  \right. \right. & \left. e^{j\tau_m^{(t+1)}} \mathbf{B}_m^H \mathbf{a}_m(\theta)  \right)^H \cdot\displaystyle\left(\hat{\mathbf{y}}_m- g^{(t+1)}_\text{opt} e^{j\tau_m^{(t+1)}} \mathbf{B}_m^H \frac{\partial \mathbf{a}_m(\theta) }{\partial \theta} \right)+\\
 &&\displaystyle \left. \left(\hat{\mathbf{y}}_m^H- g^{(t+1)}_\text{opt} e^{-j\tau_m^{(t+1)}} \frac{\partial\mathbf{a}_m(\theta)^H}{\partial \theta}  \mathbf{B}_m  \right) \cdot \left(\hat{\mathbf{y}}_m- g^{(t+1)}_\text{opt} e^{j\tau_m^{(t+1)}} \mathbf{B}_m^H \mathbf{a}_m(\theta)  \right)\right]
\end{array}
\label{gradient_calu}
	\end{equation}
	\setcounter{equation}{\value{mytempeqncnt}}
	\hrulefill
	\vspace*{4pt}
\end{figure*}
\setcounter{equation}{34}

Finally, based on the updated beam gain information $ge^{j\tau_m}$, we optimize the beam direction $\hat{\theta}^{(t+1)}$  via gradient descent method \cite{convex_optimization}.  With $g^{(t)}$ and $\tau^{(t)}_m$ fixed in (\ref{further_objective}), the gradient of beam direction is calculated as (\ref{gradient_calu}), and the update process is formulated as
\begin{equation}
\hat{\theta}^{(t+1)}=\hat{\theta}^{(t)}-\eta \nabla \epsilon\left.\right. \Big|_{\hat{\theta}^{(t)}}.
\label{angle_opt}
\end{equation}

After several iterations we can obtain the accurate beam direction to further overcome power leakage problem. The detailed process is provided in Algorithm 2 as follows.

	\begin{algorithm}[htb] 
		\normalem
		\caption{Proposed Power Leakage Compensation for Wideband THz frequency-scanning-based Tracking} 
		\label{alg2} 
		\begin{algorithmic}[1] 
			\REQUIRE Received signals $y_{l,m}$, Precoding matrix $\mathbf{f}_{l,m}$ and initial tracking result $\hat{\theta}^{(0)}$ obtained from Alg 1; Maximum iteration $T$; Update threshold $\xi$. 

			\ENSURE Accurate physical direction $\hat{\theta}$
			

			\STATE Calculate the reshaped signal vector $\hat{\mathbf{y}}_m$ and precoders $\mathbf{B}_m$ in (\ref{further_objective})
            
            \FOR{iteration num $t=0$ to $T$}
            \STATE Update the common path gain amplitude $g^{(t+1)}$ among subcarriers and timeslots via (\ref{gain_opt})
            \STATE Update frequency-dependant beam gain phase $\tau_m^{(t+1)}$ via (\ref{phase_opt})
            \STATE Compensate power leakage and update the real beam direction $\hat{\theta}^{(t+1)}$ via (\ref{angle_opt})
            \IF{$\left|[g^{(t+1)},\hat{\theta}^{(t+1)}]-[g^{(t)},\hat{\theta}^{(t)}]\right|^2<\xi$}
            \STATE Break;
            \ENDIF
            \ENDFOR
            \STATE Final beam direction $\hat{\theta}\leftarrow \hat{\theta}^{(T)}$
		\end{algorithmic}
	\end{algorithm}

Furthermore, in some low-SNR scenarios, the signal-based tracking isn't always completely reliable while prior information on UE trajectory is more helpful for beam tracking. Certain Gauss-Markov UE-mobility models can be further adopted here, and extended Kalman filter (EKF) or unscented Kalman filter (UKF) can be similarly utilized based on TTD-aided tracking. The corresponding basic work on EKF/UKF tracking can be found in \cite{spatial_rotation_UKF,tracking_Markov}. The combination of TTD-aided tracking and EKF/UKF is herein omitted since the joint tracking is simple and quite intuitive. 

\section{Searching Radius Enhancement}
In this section, we focus on the two tracking algorithms' performance comparison and take the tracking range $\alpha$ as one important criterion, which directly controls the overhead consumption. Furthermore, one critical question is whether the range in (\ref{F_B_pairing_range}) is maximum or not. Fortunately the answer is negative. An exquisite enlarged range with closed form is provided in this section \footnote{Without loss of generality, we only consider the spatial beam direction $\theta\in [0,1]$ and neglect the negative interval $[-1,0)$ in Section V and Section VI, due to its symmetry properties.}.

\subsection{Forward-mapping tracking \cite{zooming_tracking} (as comparison)}

Based on the above analysis, we have given a upper-bounded searching range (\ref{forward_range}) and obtained the beam-independent maximum range $\alpha^\text{fw}=\frac{1}{P}-\frac{Mf_d}{f_c}$. Nevertheless, another criterion should also be considered in forward-mapping tracking \cite{zooming_tracking} as comparison for more fairness. Consider one-timeslot tracking here for convenience. When the period of initial beam $\Xi_{N_\text{TTD}}(\cdot)$ in (\ref{inner_beam_property}) is larger than total searching range $2\alpha$, the side lobes of $\Xi_{N_\text{TTD}}(\cdot)$ at overall subcarriers are all located outside the searching interval $[\theta_0-\alpha,\theta_0+\alpha]$, which also means there exists no interference inside searching interval. Under this criteria, the constraint can be formulated as $T_m^{(2)}>2\alpha, \ \forall m=-M,\dots,M$, and then we have 
\begin{equation}
\alpha_\text{f-ind}^\prime<\min \ \frac{f_c}{f_m}\frac{1}{P}=\frac{f_c}{f_M P}.
\label{forward_range_2}
\end{equation}

When the former bound (\ref{forward_range}) is smaller than (\ref{forward_range_2}), i.e., $\theta_0>\frac{f_c}{f_M P}$, the tolerable searching radius increases via (\ref{forward_range_2}) but is still weaker than $\frac{1}{P}$. And it is worth noting that this bound only holds in single-timeslot scenario. When overall $L>1$ pilot timeslots are adopted with each angular-mapping fraction radius as $\alpha/L$, there exists inter-fraction interference (IFI) caused by the sidelobes of windowing beam $\Xi_P(\cdot)$, which means corresponding tracking range should be further reduced.  Otherwise, the side lobe is extremely enlarged and harmfully influences other time fractions' beam pattern, which causes the wideband tracking failure.

\subsection{Forward-Backward-Pairing Tracking (Algorithm 1)}

Similarly, based on the above basic criteria, we have obtained standard searching range as (\ref{F_B_pairing_range}), which is always larger than $\frac{1}{P}$. Next we analyze and provide further enhancement on tracking radius $\alpha$.
\begin{lemma}
The periodic sidelobes \footnote{Notice that the definitions of sidelobe and mainlobe here don't depend on the beam amplitudes but the locations in (\ref{inner_beam_property}). The exact location of $\theta_{c,m}^{(2)}$ in (\ref{inner_beam_property}) is defined as mainlobe while the peak location with periodic shifting is defined as sidelobe.} of $\Xi_P(\cdot)$ in (\ref{outer_window_property}) at subcarriers $f_{-M}$ and $f_M$ are located in different sides with respect to vertical line $\theta=\theta_0-\frac{Mf_d}{f_c}\alpha$. And the all-frequency joint beam pattern (as shown in Fig. \ref{beam_diffusion}(b)) gets global maximum gain at $\theta=\theta_0-\frac{Mf_d}{f_c}\alpha$.
\label{lemma1}
\end{lemma}

\begin{IEEEproof}
The proof is provided in Appendix A.
\end{IEEEproof}

\begin{figure*}[!t]
\centering
\includegraphics[width=0.9\linewidth]{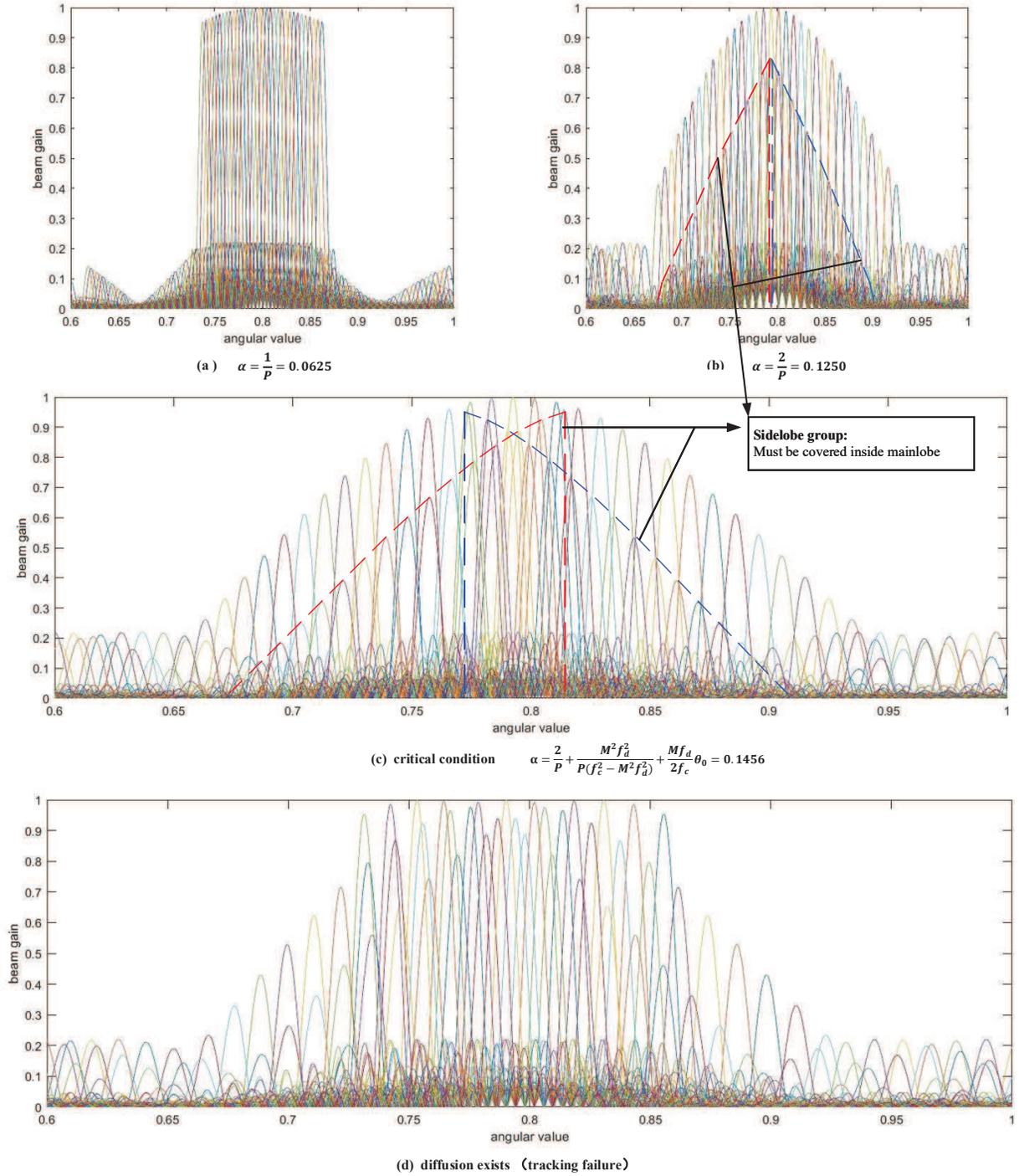}
\caption{The beam patterns of forward-backward pairing scheme under different searching ranges $\alpha$.}
\label{tracking_range}
\end{figure*}

From Lemma 1 we consider a specific situation, i.e., $\theta_{-M}^\text{Side}=\theta_M^\text{Side}=\theta_c$ as shown in Fig. \ref{tracking_range}(b). The searching range can be calculated as $\alpha=\frac{2f_c^2}{P(f_c^2-M^2f_d^2)}>\frac{2}{P}>\alpha^\text{b-f}$. Corresponding sidelobe gain  $G^\text{Side}_{-M}=G^\text{Side}_M=\Xi(\frac{Mf_d}{f_c}(\theta_0-\frac{Mf_d}{f_c}\alpha))$ is almost decreasing as central angle $\theta_0$ raises. When $\theta_0$ is small, for example, setting $\theta_0= \frac{Mf_d}{f_c}\alpha\approx 0$, the sidelobe gain can approach $G^\text{Side}_M\approx \Xi(0)=1$, which means the mainlobe group cannot effectively cover sidelobe group and cause a tracking failure. Therefore, when the direction is small enough, the maximum range is bounded by (\ref{F_B_pairing_range}) exactly. However, as central angle $\theta_0$ increases, the sidelobe gain fades as shown in Fig. \ref{diric_function}. In this way, the mainlobe's envelope can absolutely cover the sidelobes and then strongest beam can be tracked correctly via the mainlobe group, which means the searching radius $\alpha$ can be further enlarged from $\alpha^\text{f-b}$ (\ref{F_B_pairing_range}). An intuitive example is shown in Fig. \ref{tracking_range}.b, where the central angle is set as $0.8$ and the coverage of mainlobe envelope is full and acceptable.

The next question is what the accurate upper bound for the tracking range is. Consider the marginal situation, i.e., when the sidelobe group touches mainlobe envelope, as depicted in Fig. \ref{tracking_range}(c). It is worth noting that the sidelobe $\theta_M^\text{Side}$ at $f_M$ is smaller than $\theta_c$, and similarly we have $\theta_c<\theta_{-M}^\text{Side}<\theta_{-M}$. 

\begin{theorem}
With a large central direction $\theta_0\rightarrow 1$, the maximum searching radius $\alpha_\text{max}$ in forward-backward-pairing tracking scheme is bounded by 
\begin{equation}
\arraycolsep=1pt\def\arraystretch{2.2}
\begin{array}{lll}
&\displaystyle \alpha^\text{bound}&=\displaystyle \min \left\{\frac{1.620}{P}+\frac{Mf_d}{f_c}\theta_0 , \right.\\ 
&&\displaystyle \left. \frac{2}{P}+\frac{M^2f_d^2}{P(f_c^2-M^2f_d^2)}+\frac{Mf_d}{2f_c}\theta_0\right\}.
\end{array}
\label{final_upperbound}
\end{equation}
\label{theorem1}
\end{theorem}

\begin{IEEEproof}
	The proof is provided in Appendix B.
\end{IEEEproof}

From Theorem 1, we can observe that the searching range can be further extended when with large central angle situation. For some small searching angle $\theta_0\rightarrow 0$, we suggest the tracking range $\alpha$ could be set as (\ref{F_B_pairing_range})  and when $\theta_0\rightarrow 1$ we set the range as (\ref{final_upperbound}). Furthermore, according to the overall analysis above, when we consider the fixed searching radius $\alpha$ (irrelevant to angle $\theta_0$) for lower hardware/power consumption, we suggest the \textbf{global-angular-independent range} to be set as
\begin{equation} \alpha^\text{fixed}=\frac{1}{P}.
\end{equation}
Otherwise, \textbf{quasi-angular-independent} searching range can be set via certain piecewise functions for instance: 
\begin{equation}
\alpha^\text{quasi-fixed}=\left\{
    \arraycolsep=1pt\def\arraystretch{2.2}
        \begin{array}{lll}
        &\displaystyle \frac{1}{P}, \ \ &\displaystyle |\theta_0|<\theta^\text{th}_1,\\
        &\displaystyle \frac{1.620}{P}, \ \ &\displaystyle \theta^\text{th}_1<|\theta_0|<\theta^\text{th}_2\\
        &\displaystyle \frac{2}{P}\ \ (+\frac{M^2f_d^2}{Pf_1f_M})\ , \ \ &\displaystyle |\theta_0|>\theta^\text{th}_2
        \end{array}
        \right.
        \label{quasi_fixed}
\end{equation}
where $\theta^\text{th}_1$ and $\theta^\text{th}_2$ denote threshold angles for piecewise function, which is related with system equipment such as antennas $N$, TTD number $N_\text{TTD}$ and frequency configurations $M$, $f_c$, $B$ and so on. Conservatively speaking, we can uniformly divide the AoD domain at $\text{AoD}=30\degree$ and $\text{AoD}=60\degree$, i.e., set $\theta_1^\text{th}=0.5$ and $\theta_2^\text{th}=0.866$ for the quasi-fixed searching radius in (\ref{quasi_fixed}).

\section{Finite-Codebook-based Tracking}

In this section, we give a joint consideration of TTD-aided tracking and beamforming. The above analysis is based on the infinite resolution of $\psi$ and $t$ in (\ref{codeword}). However the beamforming codebook size is always countable in practice, which inspires us to evaluate the performance with finite beamforming codeword configuration.

As for beamforming module, given the DFT-based codeword pattern derived in (\ref{codeword}), the quantization for $\psi$ and $t$ are easily yielded to 
\begin{equation}
\psi^\text{BF}\in \mathbb{U}^Q_{2/N_\text{BS}}\left([-1,1]\right), \ \  t^\text{BF}\in \mathbb{U}^Q_{2/N_\text{BS}}\left([-1,1]\right),
\label{beamforming_codebook}
\end{equation}
which denotes uniform quantization for the whole angular domain $[-1,1]$ with codebook size as $N_\text{BS}$ (resolution as $\Delta t^\text{BF}=2/N_\text{BS}$). Then at each subcarrier, $N_\text{BS}$ almost-orthogonal beams are generated  to provide full communication direction coverage, just like the conventional narrowband DFT-based analog beamforming.

As for tracking module, in the forward-pairing scheme, substituting the radius constraint (\ref{forward_range}) to $t^\text{fw}$ (\ref{forward_codebook_setting}), we can get that $t^\text{fw}=\frac{Mf_d}{f_c}\frac{1}{P}\geq 1$ since $\frac{1}{P}=\frac{N_\text{TTD}}{N_\text{BS}}\geq \frac{Mf_d}{f_c}$ according to \cite[equ.25]{beamforming_TTD_delay_phase}. We can observe that the codewords in beamforming ($|t^\text{BF}|\leq 1$) and tracking ($t^\text{fw}>1$) are incompatible if we adopt the tracking radius upper bound (\ref{forward_range}), and thus the fixed radius $\alpha^\text{fixed,fw}$ should be further reduced from $\frac{1}{P}-\frac{Mf_d}{f_c}$.  From another aspect, if we utilize the beamforming codebook (\ref{beamforming_codebook}) for forward-pairing track, we should ensure that $t^\text{fw}\leq1$ and then get $\alpha^\text{fw}\leq \frac{Mf_d}{f_c}(1-\theta_0)$, which means the fixed searching radius doesn't exist at all.  

On the other hand, next we give an analysis of the proposed forward-backward-pairing scheme. We also only consider the angular interval $[0,1]$ (i.e. the backward-pairing method) and neglect $[-1,0]$ part due to its symmetry property. If we still utilize the above codebook (\ref{beamforming_codebook}) for tracking, we have $t^\text{bw}=\theta_0-\frac{f_c}{Mf_d}\alpha\in [-1,1]$, then the corresponding fixed tracking radius $\hat{\alpha}^\text{fixed}$ should be modified as
\begin{equation}
\alpha\leq\frac{Mf_d}{f_c}(1+\theta_0)\ \Rightarrow \hat{\alpha}^\text{fixed}=\frac{Mf_d}{f_c}\leq \alpha^\text{fixed}.
\label{fixed_BR_alpha}
\end{equation}
From (\ref{fixed_BR_alpha}) we can get that the fixed tracking radius still exists but decreases when initial beamforming codebook (\ref{beamforming_codebook}) is adopted. To enlarge the radius to $\alpha^\text{fixed}$ or even $\alpha^\text{quasi-fixed}$ with quantized $\psi$ and $t$, the quantization interval of $t$ should be further entended. For example, if we select the fixed maximum radius $\alpha^\text{fixed}=\frac{1}{P}$, the minimum value of $t$ in tracking is calculated as
\begin{equation}
t_\text{min}=\min_{\theta_0}\ \theta_0-\frac{f_c}{Mf_d}\alpha^\text{fixed}=-\frac{f_c}{Mf_d\cdot P}
\end{equation}
Similarly the maximum value is $t_\text{max}=\frac{f_c}{Mf_d\cdot P}$

 As far as the quantization interval is designed, we should further optimize the resolution for $t$ correspondingly. Notice that the quantized step of $\theta_0$ is $2\alpha^\text{fixed}$ in tracking, and $f_c/Mf_d\alpha^\text{fixed}\gg \alpha^\text{fixed}$. Therefore, the latter term $\frac{Mf_d}{f_c}\alpha^\text{fixed}$ is dominant for resolution design of $\psi$ while the former term $\theta_0$ is dominant for $t$ in (\ref{backward_codebook_setting}). Then we can obtain that for $t$, the resolution $\Delta t$ should satisfy that 
\begin{equation}
\Delta t^\text{Track}\leq 2\alpha^\text{fixed}=\frac{2}{P}
\label{res_track}
\end{equation}
 and therefore, with joint consideration of TTD-aided beamforming and tracking, the quantization of $t$ can be reformulated as follows via (\ref{beamforming_codebook}) and (\ref{res_track})
\begin{equation}
\arraycolsep=1pt\def\arraystretch{2.2}
\begin{array}{lll}
t\in&&\displaystyle  \mathbb{U}^Q_{2/P}\left(\left[-\frac{f_c}{Mf_d\cdot P},-1\right]\right) \  \cup \\ && \displaystyle  \mathbb{U}^Q_{2/N_\text{BS}}\left([-1,1]\right)\  \cup \  \mathbb{U}^Q_{2/P}\left(\left[1,\frac{f_c}{Mf_d\cdot P}\right]\right)
\end{array}
\end{equation}

\section{Simulation Results}
In this section, we provide some simulations to confirm the brilliant performance of our proposed forward-backward tracking scheme and power leakage compensation method. 
In the simulation, we assume BS is equipped with ULA antennas $N_\text{BS}=256$ and only consider $K=4$ single-antenna UEs for tracking and beamforming.  TTD line number and other corresponding system parameters are set as $N_\text{TTD}=16$, $P=16$ and $N_\text{RF}=K$, respectively. The carrier frequency is $f_c=100\ \text{GHz}$. Baseband semi-bandwidth and subcarrier number are set as $B=10\ \text{GHz}$ and $M=64$. As for THz wireless channel, due to the quasi-optical signal transmission from massive antennas and high frequency, we only consider LoS path for each UE, i.e., $N_k^p=1$. The path gains are randomly generated with distribution $g_p\in \mathcal{C}(1, \sigma_p^2)$. The initial physical directions of different users $\theta_k^{(0)}$ are randomly generated by the distribution following $\mathcal{U}(-1,1)$. In each frame $t$, the real physical direction changes as $\theta_k^{(t)}=\theta_k^{(t-1)}+\Delta\zeta_k^{(t)}$, where we assume the whole system has already known the last direction $\theta_k^{(t-1)}$ at $t$-th frame, and $\Delta\zeta_k^{(t)}$ denotes the statistical mobility parameter, which is always bounded by maximum bias $\zeta_\text{max}=0.2$. In another word, at $t$-th frame, we only need to track the angular interval with overall central angle $\theta_k^{(t-1)}$ and total angular length $2\zeta_\text{max}$, i.e., $[\theta_k^{(t-1)}-\zeta_\text{max}, \ \theta_k^{(t-1)}+\zeta_\text{max}]$. In each frame, we configure $L$ timeslots for TTD-based tracking and in each timeslot we search angular sub-interval with length $2\zeta_\text{max}/L$. The normalized mean square error (NMSE) and beamforming gain here are defined as follows:
\begin{equation}
\text{NMSE} = \frac{1}{K}\sum_{k=1}^{K}\mathbb{E}\left[ \frac{|\hat{\theta}-\theta_r|^2}{|\theta_r|^2} \right]
\end{equation}
\begin{equation}
\text{G}_\text{BF} = \frac{1}{KM}\sum_{k=1}^K\sum_{m=-M}^{M}|\mathbf{h}_{k,m}^H\mathbf{f}_m(\hat{\theta})|^2
\end{equation}
where $\hat{\theta}$ and $\theta_r$ are the tracking result and real channel beam direction, respectively. 

\begin{figure}[!t]
\centering
\includegraphics[width=1\linewidth]{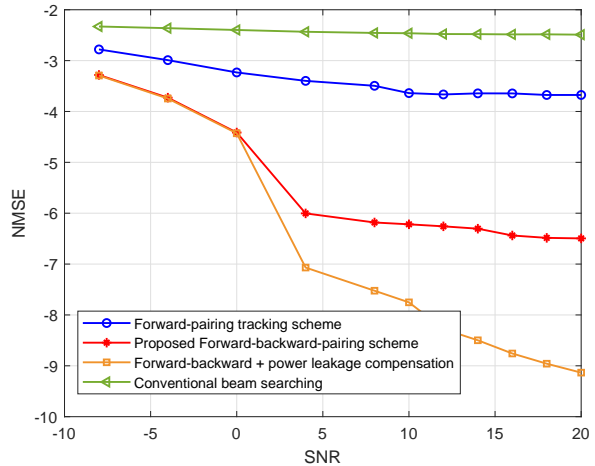}
\caption{NMSE performance comparison against SNR with pilot overhead number $L=4$ and whole searching range $\alpha=0.2$.}
\label{NMSE_SNR}
\end{figure}

\begin{figure}[!t]
\centering
\includegraphics[width=1\linewidth]{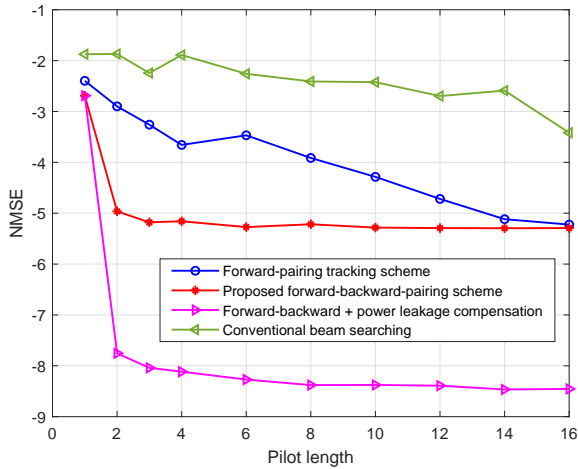}
\caption{NMSE performance comparison against pilot overhead length with SNR $10\  \text{dB}$ and whole searching range $\alpha=0.2$.}
\label{NMSE_pilot}
\end{figure}

\begin{figure*}[htbp]
\begin{center}
\subfigure[Path direction $\theta_0=0.4$ ($\text{AoD}=23.6\degree$)]{
\includegraphics[width=0.48\linewidth]{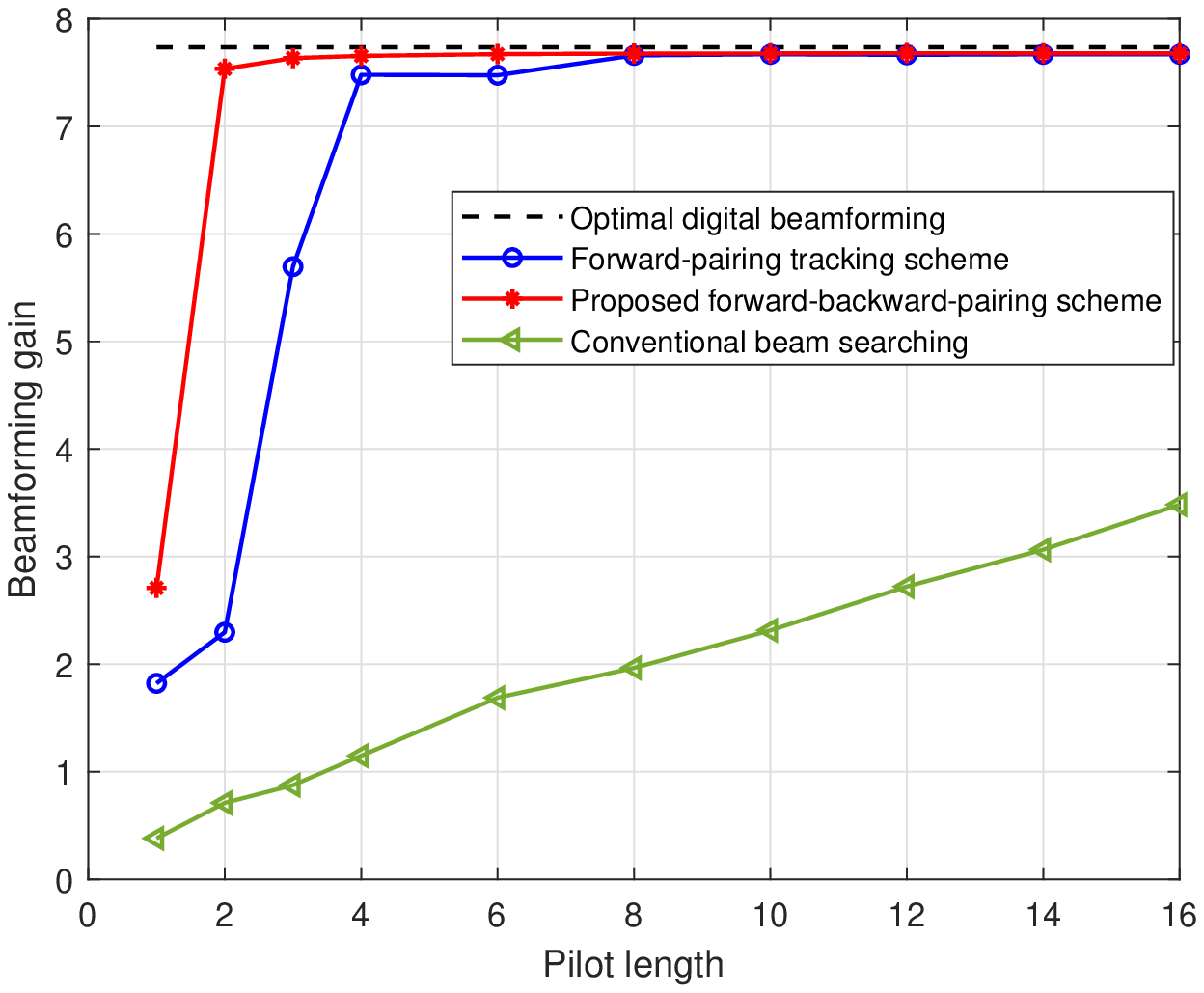}
}
\subfigure[Path direction $\theta_0=0.8$ ($\text{AoD}=53.1\degree$)]{
\includegraphics[width=0.48\linewidth]{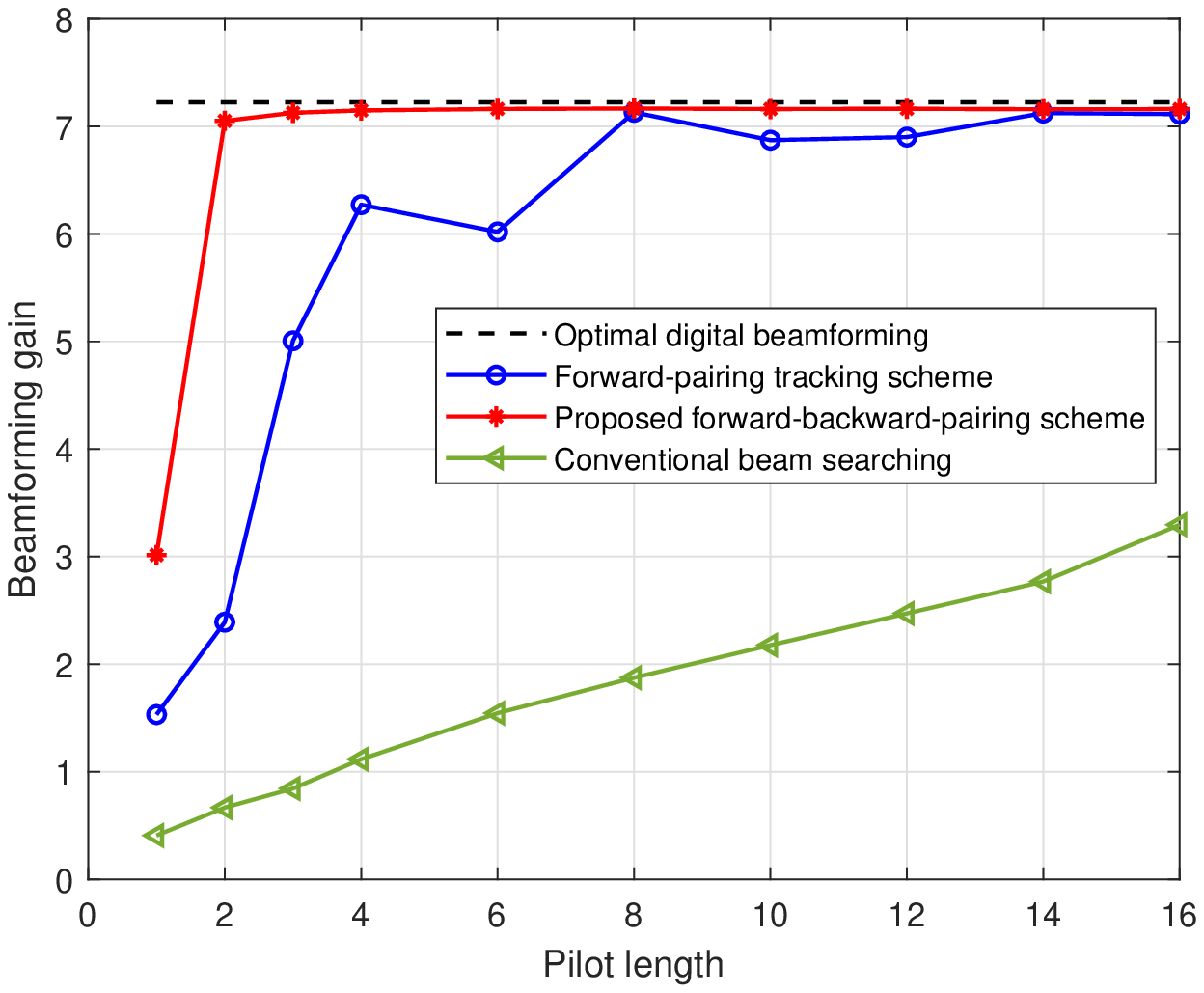}
}
\caption{Beamforming gain comparison (based on tracking results) against path direction, with SNR $10 \ \text{dB}$ and searching range $\alpha=0.2$.}
\end{center}
\label{sim_gain_pilot}
\end{figure*}

\begin{figure*}[htbp]
\begin{center}
\subfigure[Pilot length $L=2$]{
\includegraphics[width=0.48\linewidth]{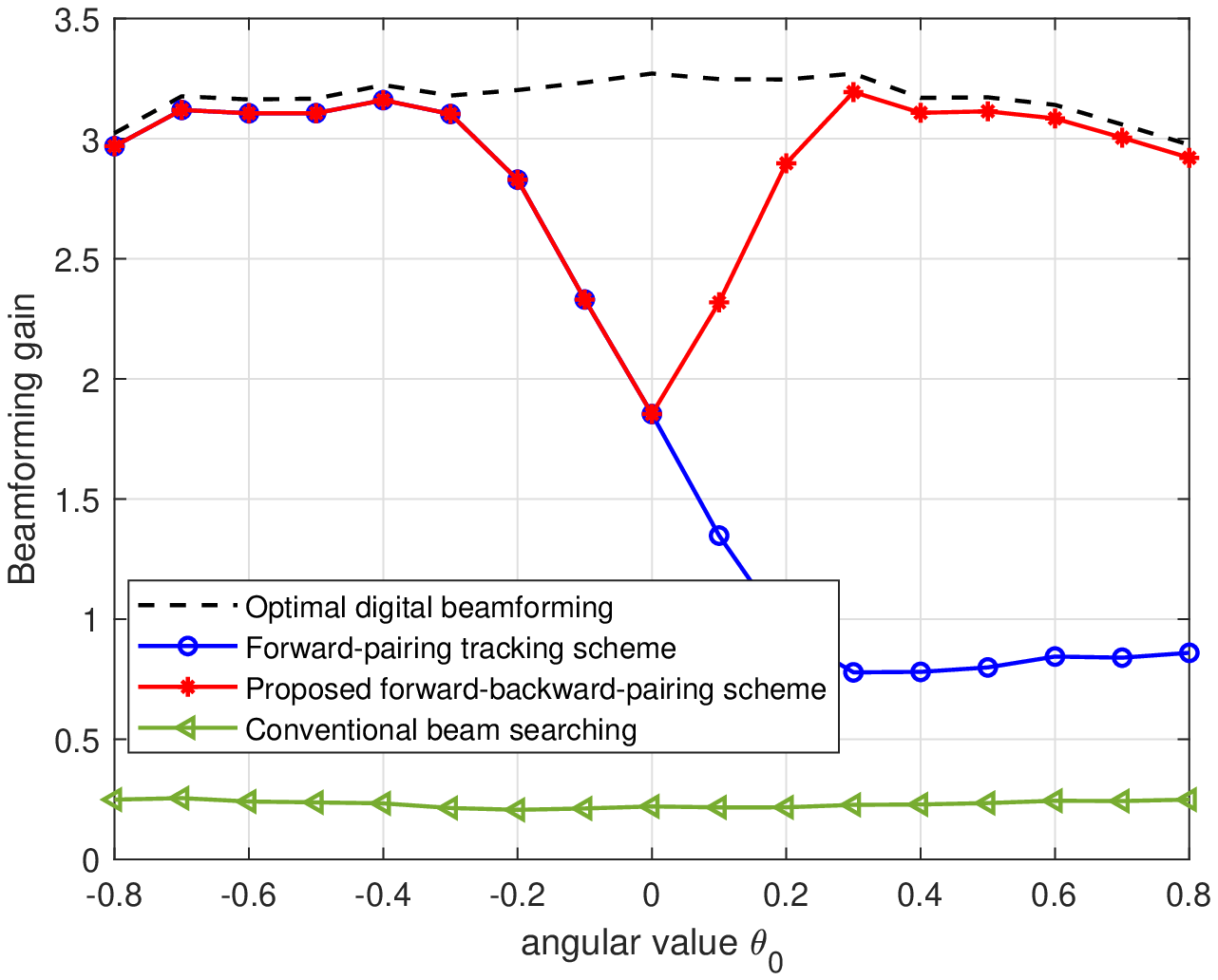}
\label{sim_gain_angle_a}
}
\subfigure[Pilot length $L=3$]{
\includegraphics[width=0.48\linewidth]{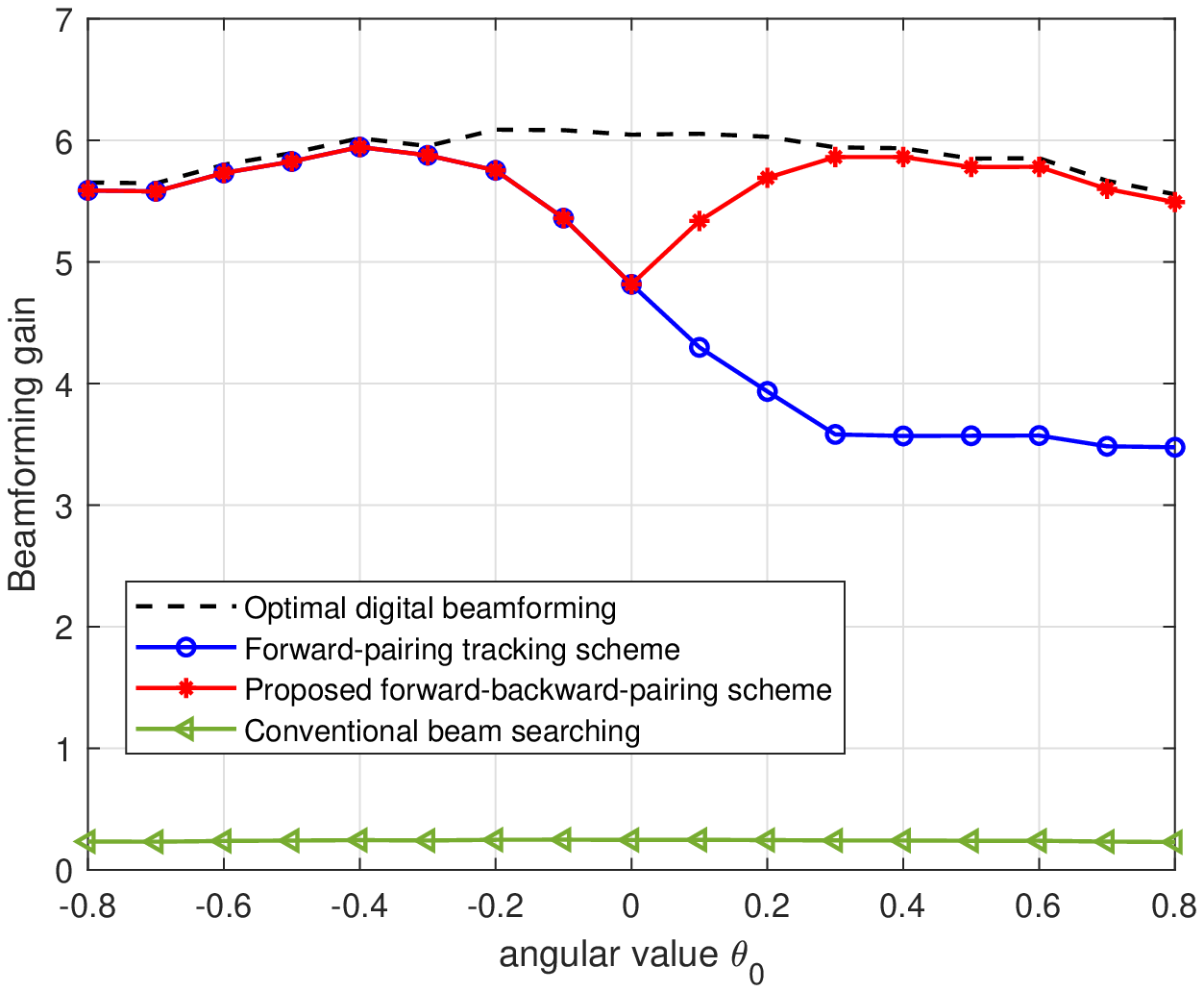}
\label{sim_gain_angle_b}
}
\caption{Beamforming gain comparison (based on tracking results) against beam direction, with SNR $10 \ \text{dB}$ and the searching range $\alpha=0.2$.}
\end{center}
\end{figure*}

Fig. 6 shows the average tracking accuracy (NMSE) by different kinds of THz tracking approaches. The method in \cite{zooming_tracking} is displayed as a benchmark, which only forwardly mapped subcarriers and beam direction. It can be seen that the NMSE of the proposed forward-backward-pairing scheme with $\alpha=0.2$ becomes lower with the increase of SNR. At high SNR, the leakage error dominates and hampers further decline of NMSE to $-6 \ \text{dB}$. Fortunately, through the additional application of leakage compensation method (Algorithm 2), the tracking error can be significantly solved as shown in Fig. 6. Besides, the conventional beam searching method is also provided here for comparison, which only sweeps one beam direction using all subcarriers in one timeslot. In Fig. 6 with pilot overhead $L=4$, without the TTD-aided simultaneous angular pairing of several subcarriers, the tracking accuracy and efficiency is deadly reduced to $-2.4 \ \text{dB}$. 

Fig. 7 presents the NMSE performance under different pilot overhead lengths $L$. The whole angular searching interval is fixed as $[0.4,0.8]$, i.e., $\theta_0=0.6$ and $\alpha=0.2$. This interval is firstly divided into $L$ fractions and then only the corresponding fraction is swept in each pilot timeslot $l=1,\dots,L$. The benchmarks here are the same with Fig. 6. With abundant pilot overhead consumption (for instant $L=16$ in Fig. 7), the method in \cite{zooming_tracking} performs quite similarly to our proposed scheme. However, with low pilot length, our proposed forward-backward-pairing scheme extremely outperforms method in \cite{zooming_tracking} with NMSE gap almost $1.7 \ \text{dB}$ at pilot length $L=6$. Besides, it is worth noting that only $2$ or $3$ pilot timeslots are needed to sweep a large angular interval and achieve quite low tracking loss through our proposed scheme.

The beamforming gains against pilot length are shown in Fig. 8(a) and 8(b). All curves here adopt the same precoding method that exactly align to the tracked beam direction, while the tracking accuracy varies among different approaches. Frequency-dependent fully-digital  precoder with perfect CSI is displayed as an upper bound for comparison. SNR is set as $10\ \text{dB}$ and the whole searching range is fixed as $\alpha=0.2$. As shown in Fig. 8, there exists a huge gap between frequency-scanning-based tracking and conventional searching-based method \cite{tracking_hierarchical}, which has been acknowledged and analyzed in previous study \cite{zooming_tracking}. To approach beamforming gain $7.5$, only $2$ pilot timeslots are needed through our proposed forward-backward-pairing tracking, while $4$ pilots should be equipped for forward-pairing method. Almost $50\%$ pilot overhead can be further saved for wideband THz beam tracking.

Next in Fig. 9, we elaborate the beamforming gain under different UE locations, i.e., at different physical angular directions $\theta_r$. Interestingly, it can be observed that our proposed scheme's performance is symmetric with respect to vertical line $\theta_0=0$, while the tracking accuracy of \cite{zooming_tracking} rapidly drops off when the UE direction $\theta_r>-0.2$. From the two figures we can get that the two tracking schemes obtain the same beamforming gain in the negative angular zone,  and when $\theta_r>0$, our proposed scheme still works well but \cite{zooming_tracking} may carry tracking failure and consequently consumes larger pilot overhead. As for our proposed scheme, it can be seen that the worst performance appears when the physical UE direction is $\theta_r=0$. This can be explained by the maximum searching range in (\ref{F_B_pairing_range}) and (\ref{final_upperbound}). Under our simulation parameter settings, when UE direction angle is small ($|\theta_r|\rightarrow 0$), the maximum searching range in one timeslot is bounded by $1/P=0.0625$. To fully sweep the whole interval with radius $\alpha=0.2$, more than $\frac{\alpha}{1/P}=3.2$ pilot timeslots are needed. As shown in Fig. \ref{sim_gain_angle_a} and Fig. \ref{sim_gain_angle_b}, when pilot length is enlarged from $L=2$ to $3$, the beamforming gain at direction $\theta_r=0$ is drastically ameliorated. On the other hand, for the directions aside ($|\theta_r|\rightarrow 1$), since the searching range bound in one timeslot is calculated from (\ref{final_upperbound}) as
\begin{equation}
\frac{2}{P}+\frac{M^2f_d^2}{P(f_c^2-M^2f_d^2)}+\frac{Mf_d}{2f_c}\approx 0.1502,
\end{equation}it is sufficient to configure $2$ pilot overhead timeslots for tracking as shown in Fig. \ref{sim_gain_angle_a}. 


\section{Conclusion}
In this paper, based on TTD-aided hybrid precoding structure, we give an enhanced frequency-scanning-based tracking scheme, where we flexibly control the angular coverage from all squint beams over the whole bandwidth with two different pairing policies, to simultaneously search multiple physical directions in one timeslot. Closed-form searching radius, parameter configuration and interference are theoretically analyzed. Pilot overhead is further reduced and tracking robustness is also largely improved via our proposed scheme. Besides, we provide coupled codebook design and codeword resolution design for TTDs and PSs, with joint consideration of both beamforming and tracking. Analytical and numerical results demonstrate the superiority of the new coupled TTD-aided codebook from both tracking error and beamforming performance. Overall, the tracking accuracy and beamforming gain in wideband THz system are largely enhanced via our proposed codebooks and corresponding algorithms.

\appendices
\section{Proof of Lemma \ref{lemma1}}\label{AppendixA}

\begin{IEEEproof}
	Substituting (\ref{inner_beam_property}) and (\ref{backward_codebook_setting}) into the beam pattern, the periodic sidelobes' angular locations of $\Xi_\text{P}(\frac{f_m}{f_c}\theta-\psi)$ at $f_{-M}$, $f_M$ and the mainlobe's central location at $f_c=f_0$ are respectively written as 
	\begin{equation}
	\left\{
	\arraycolsep=0.5pt\def\arraystretch{1.8}
	\begin{array}{ccc}
	\theta_{-M}^\text{Side}&=&\theta_0+\alpha-\frac{2f_c}{Pf_1}\\
	\theta_{M}^\text{Side}&=&\theta_0-\alpha+\frac{2f_c}{Pf_M}\\
	\theta_c&=&\theta_0-\frac{Mf_d}{f_c}\alpha
	\end{array}
	\right. .
	\label{periodic_sidelobe}
	\end{equation}
	
	Reweight the top two angles with normalized factors $\frac{f_{-M}}{f_{-M}+f_M}$ and $\frac{f_{M}}{f_{-M}+f_M}$, then we have
	\begin{equation}
	\frac{f_{-M}}{f_{-M}+f_M} \theta_{-M}^\text{Side} + \frac{f_{M}}{f_{-M}+f_M} \theta_M^\text{Side} = \theta_0-\frac{Mf_d}{f_c}\alpha=\theta_c.
	\end{equation}and we can further draw the conclusion that $\theta_{-M}^\text{Side}$ and $\theta_M^\text{Side}$ are alwaye located in different sides of $\theta_c$. 
	
	Besides, substitute (\ref{backward_codebook_setting}) and $\theta=\theta_c$ back into (\ref{gain_formulation}), and we can calculate the beam gain at $f_c$ as $\mathcal{G}(f_c,\theta_c)=\Xi_P(0)\cdot\Xi_{N_\text{TTD}}(0)=1$, i.e., the upper bound of normalized Dirichlet function. Then we finish the proof for Lemma 1.  
\end{IEEEproof}

\section{Proof of Theorem \ref{theorem1}}\label{AppendixB}

\begin{IEEEproof}
	The tracking range is limited by two aspects: intra-fraction interference and inter-fraction interference. We will provide analyses of the two aspects, respectively.
	\begin{itemize}
		\item Inter-fraction interference:
	\end{itemize}
	
	Among several timeslots, the interference gain is mainly controlled by sidelobes of window function $\Xi_P(\cdot)$, while the interference angular location is limited by initial beam $\Xi_\text{TTD}(\cdot)$ in (\ref{gain_formulation}). To confirm that the $(l-1)$-th fraction's window sidelobes couldn't hold deadly interference on $l$-th angular fraction's tracking, the array gains of all subcarriers should be set larger than the maximum sidelobe gain $G_{\text{win},P}^\text{Side}$ in windowing beam $\Xi_P(\cdot)$, i.e., 
	\begin{equation}
	\arraycolsep=1pt\def\arraystretch{2.5}
	\begin{array}{lll}
	&&\displaystyle \Xi_{P}\left(\frac{f_m}{f_c}\theta_m-\psi^\text{bw}\right)>G_{\text{win},P}^\text{Side},\ \  \forall m\\
	\overset{(a)}{\Rightarrow} &&\displaystyle  \Xi_{P}\left(\frac{Mf_d}{f_c}\theta_0-\alpha \right)>G_{\text{win},P}^\text{Side}\\
	\overset{(b)}{\Rightarrow}&&\displaystyle \alpha<\Xi_P^{-1}(G_{\text{win},P}^\text{Side})+\frac{Mf_d}{f_c}\theta_0\\
	&& \displaystyle\ \ \ \overset{(c)}{\approx} \frac{1.620}{P}+\frac{Mf_d}{f_c}\theta_0,
	\end{array}
	\label{inter_fraction_range}
	\end{equation}
	where $G_{\text{win},P}^\text{Side}$ is nearly a constant value only related to $P$ since $f_m/f_c$ can be approximated to $1$ in (\ref{inner_beam_property}), and (a) is because the array gains are minimized at peripheral subcarriers $f_M$ and $f_{-M}$. (b) is derived due to the monotone increasing property of Dirichlet function $\Xi_P(x)$ inside $[-\frac{2}{P}, 0]$. And $x=\Xi^{-1}(a)$ denotes the angular location $x>0$ in the mainlobe of $\Xi(\cdot)$ with relationship $a=\Xi(x)$. Besides, (c) is approximated due to the similarity between Dirichlet function $\Xi_P(x)$ and sinc function $\text{sinc}(\frac{P}{2}x)$.
	
	\begin{itemize}
		\item Intra-fraction interference: 
	\end{itemize}
	
	For each angular fraction, we only consider intra-fraction interference inside windowing mainlobe. Notice that in the window $\Xi_P(\cdot)$'s mainlobe there exist two beams corresponding to each subcarrier as analyzed in Section III. One is utilized here positively for tracking but the other unfortunately degrades other subcarriers' tracking performance, just as shown in Fig. \ref{tracking_range}. We name the harmful beam sets along all subcarriers (circled in Fig. \ref{tracking_range}) as sidelobe groups.
	
	The marginal location of sidelobe groups (corresponding to maximum sidelobe gains),  at subcarriers $f_M$ and $f_{-M}$, are formulated in (\ref{periodic_sidelobe}). First, we only consider one side $f_M$. Substituting (\ref{periodic_sidelobe}) to (\ref{16}), we rewrite the relationship between angular location $\theta_M^\text{Side}$ and the corresponding mainlobe frequency $f^\prime$ as follows:
	\begin{equation}
	\theta_M^\text{Side}=\frac{f_c}{f^\prime}\psi^\text{bw}+\frac{f^\prime-f_c}{f^\prime}t^\text{bw}.
	\label{side_location_relation}
	\end{equation}
	Substituting $\psi^\text{bw}$ and $t^\text{bw}$ (\ref{backward_codebook_setting}) to (\ref{side_location_relation}) and we can solve the mainlobe frequency $f^\prime$ as
	\begin{equation}
	f^\prime=\frac{Pf_{-M}f_M^2}{Pf_{-M}f_M+\frac{2Mf_df_c}{\alpha}}.
	\label{sidelobe_frequency}
	\end{equation}
	$f^\prime$ here denotes corresponding subcarrier that form strongest beam at location $\theta_M^\text{Side}$, via configuration of $\psi^\text{bw}$ and $t^\text{bw}$ in (\ref{backward_codebook_setting}).
	Due to the fact $\theta_M^\text{Side}<\theta_c$, we can project it back to subcarriers and easily obtain $f^\prime>f_c$.
	
	It is not difficult to know, when the mainlobe envelope fully covers sidelobe groups, frequency-scanning-based tracking can always succeed (regardless of noise). Next, consider the critical condition as shown in Fig. \ref{tracking_range}c, the corresponding gains should be set as $\mathcal{G}(f^\prime,\theta_M^\text{Side})>\mathcal{G}(f_M,\theta_M^\text{Side})$. And according to the gain domination of windowing beam $\Xi_P(\cdot)$ in (\ref{gain_formulation}), it is equivalent to
	\begin{equation}
	\Xi_P\left(\frac{f^\prime}{f_c}\theta_M^\text{Side}-\psi^\text{bw}\right)>\Xi_P\left(\frac{f_M}{f_c}\theta_M^\text{Side}-\psi^\text{bw}\right).
	\label{inequality_gain}
	\end{equation}
	According to Lemma 1
	, $\theta_M$ and $\theta_M^\text{Side}$ are located in different sides with respect of $\theta_{c,M}^{(1)}$ (\ref{outer_window_property}), which means $\theta_M^\text{Side}>\frac{f_c}{f_M}\psi^\text{bw}$. Similarly we have $\theta_M^\text{Side}<\frac{f_c}{f^\prime}\psi^\text{bw}$. According to the monotonicity and duality of Dirichlet function $\Xi(\cdot)$, inequality (\ref{inequality_gain}) can be rewritten as 
	\begin{equation}
	\psi^\text{bw}-\frac{f^\prime}{f_c}\theta_M^\text{Side}>\frac{f_M}{f_c}\theta_M^\text{Side}-\psi^\text{bw}.
	\label{relation_envelope}
	\end{equation}
	
	Substituting (\ref{backward_codebook_setting}), (\ref{periodic_sidelobe}) to (\ref{relation_envelope}) yields that
	\begin{equation}
	2\alpha-Mf_d(\frac{1}{f^\prime}+\frac{1}{f_M})\alpha-\frac{4f_c}{Pf_M}<(2-\frac{f_c}{f^\prime}-\frac{f_c}{f_M})\theta_0.
	\label{further_inequality}
	\end{equation}
	And here (\ref{further_inequality}) is a quadratic inequality for searching radius $\alpha$, which means the solution is not that intuitive. Due to the fact $f^\prime<f_c$, we approximate the right side of (\ref{further_inequality}) and  then yield that 
	\begin{equation}
	2\alpha-Mf_d(\frac{1}{f^\prime}+\frac{1}{f_M})\alpha-\frac{4f_c}{Pf_M}<(1-\frac{f_c}{f_M})\theta_0
	\label{relaxed_inequality}
	\end{equation}
	
	Then substituting $f^\prime$ (\ref{sidelobe_frequency}) to (\ref{relaxed_inequality}), we have
	\begin{equation}
	\arraycolsep=1pt\def\arraystretch{2.5}
	\begin{array}{lll}
	&&\displaystyle (2-\frac{2f_cM^2f_d^2}{Pf_1f_M^2\alpha}-\frac{2Mf_d}{f_M})\alpha<\frac{4f_c}{Pf_M}+\frac{Mf_d}{f_M}\theta_0\\
	&\displaystyle \Rightarrow\ \ &\displaystyle \frac{2f_c}{f_M}\alpha<\frac{4f_c}{Pf_M}+\frac{2f_cM^2f_d^2}{Pf_1f_M^2}+\frac{Mf_d}{f_M}\theta_0\\
	&\displaystyle \Rightarrow\ \ &\displaystyle \alpha<\frac{2}{P}+\frac{M^2f_d^2}{Pf_1f_M}+\frac{Mf_d}{2f_c}\theta_0.
	\end{array}
	\label{intra_fraction_range}
	\end{equation}
	
	Similarly, we consider $f_{-M}$ side and can derive the same closed-form upper bound in (\ref{intra_fraction_range}). With joint consideration of intra-fraction (\ref{intra_fraction_range}) and inter-fraction interference (\ref{inter_fraction_range}) we can finish this proof. 
\end{IEEEproof}

	\ifCLASSOPTIONcaptionsoff
	\newpage
	\fi
	
	\bibliographystyle{ieeetr}
	\normalem
	\bibliography{bibliography.bib}
    
\end{document}